\documentclass[preprints,review,accept,oneauthors,pdftex]{Definitions/mdpi} 

\firstpage{1} 
\pubvolume{1}
\issuenum{1}
\articlenumber{0}
\pubyear{2022}
\copyrightyear{2022}
\datereceived{} 
\dateaccepted{} 
\datepublished{} 
\hreflink{https://doi.org/} 

\def\approxprop{%
  \def\p{%
    \setbox0=\vbox{\hbox{$\propto$}}%
    \ht0=0.7ex \box0 }%
  \def\s{%
    \vbox{\hbox{$\sim$}}%
  }%
  \mathrel{\raisebox{0.7ex}{%
      \mbox{$\underset{\s}{\p}$}%
    }}%
}

\Title{Eclipsing Binaries in Dynamically Interacting Close, Multiple Systems}

\TitleCitation{Eclipsing Binaries in Dynamically Interacting Close, Multiple Systems}

\Author{Tam\'as Borkovits $^{1,2}$\orcidA{}}

\AuthorNames{Tam\'as Borkovits}

\AuthorCitation{Borkovits, T.}

\address{%
$^{1}$ \quad Baja Astronomical Observatory of University of Szeged, H-6500 Baja, Szegedi \'ut, Kt. 766, Hungary; borko@electra.bajaobs.hu\\
$^{2}$ \quad Konkoly Observatory, Research Centre for Astronomy and Earth Sciences,  H-1121 Budapest, Konkoly Thege Mikl\'os \'ut 15-17, Hungary}

\corres{Correspondence: borko@electra.bajaobs.hu}

\abstract{Close, compact, hierarchical, multiple stellar systems, i.e., multiples having an outer orbital period from months to a few years, comprise a small, but continuously growing group of the triple and multiple star zoo. Many of them consist of at least one eclipsing pair of stars and, therefore, exhibit readily observable short-term dynamical interactions among the components. Thus, their dynamical and astrophysical properties can be explored with high precision. In this paper we present an overview of the history of the search for additional components around eclipsing binaries from the first serendipitous discoveries to more systematic recent studies. We describe the different observational detection methods and discuss their connections to the different kinds of astrophysical and dynamical information that can be mined from the different datasets. Moreover, the connection amongst the observable phenomena and the long-term dynamics of such systems is also discussed.}


\begin{document}

\section{Introduction}

In the last years of the 19th century, in a series of papers Chandler \cite{chandler888,chandler892,chandler893,chandler901} investigated the nature of some `inequalities' in the eclipsing periods of Algol (noticed first by Argelander~\cite{argelander855}) and some other eclipsing binaries (EB).  He concluded that: \textit{`Whatever the nature of the pure elliptic motion, the times of conjunction must follow each other at exactly equal intervals, unless, {\rm(a)} one or both components are not spherical; or, unless, {\rm(b)}, there is at least one more body in the system.'} And, for the first time, he also calculated the formulae of the astrometric orbit of an EB around the common center of mass of the triple star system, and the variations in the expected eclipse times (ETVs) due to the light-travel time effect (LTTE).

Chandler's two  effects that can lead to apparent period changes were too restrictive in that several other effects may also cause ETVs in an EB (for a recent review of the origins of the true and apparent period variations see Rovithis-Livaniou \cite{rovithis-livaniou20}). On the other hand, Chandler was correct in the sense that the most common and extensively studied periodic eclipse timing variations over the last 100--120 years are indeed nicely related to the two effects he envisioned. Namely, before the era of the exoplanet hunting space telescopes (see below in Sect.~\ref{sec:detection_methods}), whenever variable star observers detected quasi-sinusoidal variations in the observed versus expected mid-eclipse times of any EB, then, if the variations in the primary and the secondary eclipses were of opposite signs, they could assume with great certainty that their origin was dominated by apsidal motion caused by the non-spherical mass distribution of the stars. On the other hand, if the primary and secondary eclipses exhibited similar variations in both amplitudes and sign, it was usually attributed to LTTE caused by a third star.  Of course, what we have learned over the last decades is that things are not nearly so simple. First, the apsidal motion in eccentric binaries may arise not only from {\em (i)} the non-spherical (tidally distorted) mass distributions of the stars (see, e.g., Tisserand \cite{tisserand895}; Cowling \cite{cowling938}; Sterne \cite{sterne939} and, from non-aligned stellar rotations, Shakura \cite{shakura985} and Heged\"us \& Nuspl \cite{hegedusnuspl986}); but also, due to {\em (ii)} relativistic effects (see, e.g., Weinberg \cite{weinberg972}); and {\em (iii)} perturbations due to one or more additional stellar components (Borkovits et al. \cite{borkovitsetal07,borkovitsetal19a}). Second, the observable effects of a third (or even fourth, or more) body orbiting around the EB may not be restricted only to {\em (i)} the classical LTTE (see, e.g., Irwin \cite{irwin952,irwin959}), but also to the {\em (ii)} perturbations to the Keplerian motion of the EB due to the gravitational interactions of its components with the additional (close) companions (see, e.g., Borkovits et al. \cite{borkovitsetal15}).

Moreover, although those irregular period variations that Chandler claimed to be effects of a third body actually have a different origin, his hypothesis was correct in that sense that Algol really does have a quite short outer period ($P_\mathrm{out}=680$\,d) third stellar companion (discovered spectroscopically by Belopolski \cite{belopolski906,belopolski911} and Curtiss \cite{curtiss908}). The third body therefore makes this emblematic system, besides being the first known EB, also the first known compact hierarchical triple stellar system (CHT).

Regarding compact hierarchical triple or multiple systems, Tokovinin \cite{tokovinin21} defines them as triple systems, or triple subsystems of higher order hierarchies with $P_\mathrm{out}\leq1000$\,days. He finds, that the number of these systems are small relative to wider multiple systems. According to Tokovinin \cite{tokovinin14b} their rareness is most obvious in the volume-limited sample of FG-type stars. In his recent review, Tokovinin \cite{tokovinin21} claims that the formation scenario(s) of CHTs may differ from that of the wider triples and multiples. Despite this, in what follows we do not consider the value of $P_\mathrm{out}=1000$\,d to be a strict limit. Instead, from our observational point of view, it is at least partly an intuitive or, practical limit. For example, for $P_\mathrm{out}=1000\,$d the probability of outer or, third-body, eclipse\footnote{While the probability of the eclipses (being either total or grazing ones) in an ordinary binary depends on the sum of the fractional radii of the two stars, calculating the same quantity for third-body eclipses, one may use the half of the ratio of the inner to outer semi-major axes as an effective fractional radius of the inner binary (see e.g., Appendix A of Borkovits et al.~\cite{borkovitsetal13}).} in triple systems consisting of three normal stars drops to $\sim0.5\%$ (see, e.g., Borkovits et al.~\cite{borkovitsetal20a}). Thus, it is hardly surprising that all but one of the triply eclipsing triple systems (see below, in Sect.~\ref{sec:triply_eclipsers}) with known outer periods have $P_\mathrm{out}<1000$\,d. But, on the other hand, we will not drop from our discussion the only known longer outer period triply eclipsing triple star. Moreover, the shorter the outer period, the more rapidly the manifestations of the effects of the third body on the motion of the inner binary occur, which is obviously a great advantage in the detection and the investigation of such systems. 

Amongst CHTs the most interesting systems are those where the dynamical interactions, mainly the third-body perturbations, are large enough for enhanced detectability as well as for informative follow-up investigations. In the following we will refer these as `tight systems'\footnote{To be clear we are defining CHT's as triples with $P_{\rm out} \lesssim 1000$\,d, i.e., short outer periods amenable to easy detection and follow-up studies, whereas `tight' indicates a small ratio of $P_{\rm out}/P_{\rm in}$, i.e., small enough for significant third-body perturbations. Of course, most ideal is the situation where a `compact' system is also `tight'.}. Tight triples (or, multiples) are not necessarily CHTs. For example, the recently known tightest triple star system is LHS~1070 (Xia et al.~\cite{xiaetal19}), which consists of three very low mass red dwarfs with inner and outer periods of 18.2 and 99 years, respectively, i.e., it has a period ratio of $P_\mathrm{out}/P_\mathrm{in}=5.4$, which is close to the dynamical stability limit of Mardling \& Aarseth \cite{mardlingaarseth01}. Naturally, in such a tight system one can expect strong perturbations and, therefore, variations in the orbital elements.  However,  the timescale of such variations ranges from decades to centuries (see below, in Sect.~\ref{Sect:dynamics}) and, therefore, from the point of view of any human observers, it is not particularly exciting. Amongst the most auspicious systems, i.e. tight CHTs in which the inner pairs are also eclipsing binaries (ECHTs) the record-holder is KIC~7668648 with $P_{\rm out}/P_{\rm bin}\sim7.4$, which ultimately also turned out to be a triply eclipsing triple system (Borkovits et al. \cite{borkovitsetal15}; Orosz \cite{orosz15}). At the time of this writing the shortest known outer period system (with $P_\mathrm{out}=33.03$\,d) is $\lambda$~Tau (Ebbighausen \& Struve~\cite{ebbighausenstruve956}) which is an almost equally tight ECHT with $P_\mathrm{out}/P_\mathrm{in}=8.3$ as KIC 7668648.

The exceptional importance of (E)CHTs can be summarized as follows:

\begin{itemize}
\item{} The short-term perturbations (with characteristic periods of the same order as the period of the outer orbit) in a hierarchical system are driven by the same physical and dynamical quantities which determine the long-term dynamical evolution through secular perturbations (Harrington~\cite{harrington968}; Naoz~\cite{naoz16}). Thus, proper analysis of the short-term photodynamical properties and behaviour of compact multiple systems allows all key stellar and orbital parameters, as well as the dynamical evolution of the orbits, to be determined with high accuracy. 
\item{} The formation of young, unevolved, close binary stars with short periods (of a few days or less) cannot be explained without the assumption of some effective orbit-shrinking mechanism. Different mechanisms have been suggested in the literature (see below for details), however most of them require an initial gravitational interaction with additional bodies (being either bound or unbound companion stars).
\item{} Stars in close binaries evolve in markedly different ways than their single counterparts. In multiple systems the evolutionary scenarios may be even more different and variegated (Toonen et al.~\cite{toonenetal16,toonenetal20,toonenetal21}). For example, it is hypothesized that a significant fraction of the common envelope binaries (the supposed progenitors of cataclysmic variables), type Ia supernovae (Maoz et al.~\cite{maozetal14}), and contact binaries (of the W UMa type) might be formed through some interactions with a third body (Eggleton~\cite{eggleton06}; Paczy\'nski et al.~\cite{paczynskietal06}; Csizmadia et al.~\cite{csizmadiaetal07}).
\item{} The existence of some subgroups of close binary and multiple systems containing one or more compact objects (neutron stars, black holes), the progenitors of which went through supernova explosion, is hardly understandable without the initial multiplicity of such systems. For example, Podsiadlowski et al.~\cite{podsiadlowskietal03} have pointed out that, in contrast to their more massive counterparts, black hole X-ray binaries with low-mass donor stars cannot form through a previous common envelope phase of the binary star. Therefore the question of their origin remained open for a long time, while Naoz et al.~\cite{naozetal16} have shown that black-hole -- low-mass X-ray binaries can form through the dynamical interactions with a third body. Similarly, some peculiar binary pulsars have been discovered during the last few decades (see, e.g. Champion et al.~\cite{championetal08}), which probably have a triple system origin (see e.g., Freire et al.~\cite{freireetal11}; Portegies Zwart et al.~\cite{portegieszwartetal11}; Pijloo et al.~\cite{pijlooetal12}). Moreover, the discovery of the triply degenerate hierarchical triple system PSR J0337+1715, a millisecond pulsar with two orbiting white dwarfs naturally requires the consideration of triple-star evolutionary scenarios in order to understand its formation process (Tauris \& van den Heuvel~\cite{taurisvandenheuvel14}). 
\item{} Connected to the previous item, hierarchical triples may play an important role in the formation and also in the merger of black hole binaries. For example, Rodriguez \& Antonini~\cite{rodriguezantonini18} have shown that the Lidov-Kozai mechanism (see later, in Sect.~\ref{sec:apsenode}) may explain the frequency of some special classes of the observed black hole mergers. We note also, that recently Deme et al.~\cite{demeetal20} found that in some special cases, third-body interactions amongst intermediate and supermassive black holes would also be detectable with the future space-borne gravitational wave observatory LISA\footnote{\url{https://lisa.nasa.gov/}} (see, e.g., Amaro-Seoane et al.~\cite{amaro-seoaneetal17}). Naturally, on the other hand, these triple systems are very far from the category of CHTs and, therefore, in what follows, we do not deal with them. 
\item{} Last, but not least, we refer to circumbinary planets which, from a dynamical point of view, should also be considered as hierarchical triple systems. The existence of such systems challenges our planetary formation theories. It is unclear how these planets form and evolve; it is also not clear where they form and how they migrate to their present position (e.g. Pierens \& Nelson~\cite{pierensnelson13}; Marzari et al.~\cite{marzarietal13}; Meschiari~\cite{meschiari12}; Lines et al.~\cite{linesetal14,linesetal15,linesetal16}). The role of dynamical interactions in such systems as well as their stability have been studied intensively over the past few years (see the review of Marzari \& Thebault~\cite{marzarithebault19}, and further references therein). Moreover, we note that in addition to these theoretical difficulties, that there are only limited observational clues to understanding the circumbinary planet populations because the observations suffer from many selection effects as well as the observational window function which prevent us from detecting more circumbinary planets and from understanding the observational biases (see, e.g., Klagyivik et al.~\cite{klagyiviketal17}). 
\end{itemize}

All questions regarding the formation scenarios of compact hierarchical systems were excellently reviewed most recently by Tokovinin~\cite{tokovinin21}, while in the regard to their more or less exotic final states we refer the reader to the recent papers of Toonen et al.~\cite{toonenetal16,toonenetal20,toonenetal21}.  These works save us from going through the detailed discussions of these fields here. Therefore, instead of questions about the birth and final fate of such systems or, in other words, instead of the perspectives of the past and the future, we concentrate here on the present, and discuss the practical, observational questions and aspects of CHTs.  Specifically, we focus on the methods of their observational detection and the information content that can be mined from sophisticated, complex analyses of the observational material. In this regard, first we briefly discuss in Sect.~\ref{Sect:dynamics} some aspects of the dynamics of hierarchical close triple and multiple stellar systems. We do this at a level that is relevant for our main topic of the observations and analyses of eclipsing binaries in dynamically interacting close, multiple systems.  This is reviewed and discussed in detail in Sects.~\ref{sec:detection_methods} and \ref{sec:discussion}.


\section{Dynamics of CHTs}
\label{Sect:dynamics}

Multiple stellar systems almost exclusively exhibit hierarchical configurations. Restricting ourselves to triple stars\footnote{Note, higher order multiple stars can be built up from triple subsystems.}, `hierarchical' means that one of the three distances which can be formed among the three constituent stars remains substantially smaller (from a factor of $\sim4$ to several orders of magnitude) than the other two distances during at least a large portion of the life-time of the system. In such cases the motions of the three stars can be more or less well approximated by two 2-body (or Keplerian) motions. Therefore, this problem can be discussed in the framework of the (perturbed) motion of two binaries: an `inner', or close binary formed by the two closer members, and an `outer', or wide binary consisting of the more distant third star, and the center of mass (CM) of the inner binary. Then, the usual sets of orbital elements can be defined for both orbits, and the time-dependent variations of these elements describe the orbital behavior. 

The hierarchical or, stellar third-body, problem was investigated for the first time by Slavenas \cite{slavenas927a,slavenas927b} and Brown \cite{brown936a,brown936b,brown936c}, who departed from his Lunar theory (Earth and Moon, as a binary system, forms a tight, compact triple with the Sun). Brown \cite{brown936b} found that the perturbations in hierarchical triple stellar systems can be divided into three different classes, as

\begin{itemize}
\item{{\it short-period perturbations,} for which the typical period is on the order of $P_\mathrm{in}$, and the amplitude is related to $(P_\mathrm{in}/P_\mathrm{out})^2$,}
\item{{\it long-period perturbations,} with a characteristic period of $P_\mathrm{out}$, and amplitude of $P_\mathrm{in}/P_\mathrm{out}$ and,}
\item{{\it apse-node terms,} having a period about $P_\mathrm{out}^2/P_\mathrm{in}$, and an amplitude that may reach unity.}
\end{itemize}
Later, Harrington \cite{harrington968,harrington969} generalized these early works for arbitrary values of mutual inclinations and outer eccentricities and, gave third-order (octupole) solutions for the apse-node time-scale variations of the orbital elements, with the application of the double-averaging method of von Zeipel \cite{vonzeipel916a,vonzeipel916b,vonzeipel917a,vonzeipel917b}. Note, according to the more convenient nomenclature used generally in celestial mechanics, the periodic perturbations with periods on the orders of both $P_\mathrm{in}$ and $P_\mathrm{out}$ are termed `short period', while perturbations with much longer periods (i.e., the apse-node terms of Brown \cite{brown936b}) are named as `long period perturbations'. (Harrington \cite{harrington968,harrington969} himself used this latter terminology.) For our purposes, however, Brown's classification fits better and, therefore, we use his categories. 

From an observational point of view, the short period perturbations even in the tightest CHTs, generally remain below the detectability limit. Therefore, in what follows, we do not consider them. Turning to the other two kinds of perturbations, an interesting difference from an observational point of view is that, while the amplitudes of the long period perturbations depend on the tightness of the triple system through the ratio of the periods (or, equivalently, the ratio of the semi-major axes), for the apse-node terms this dependence disappears from the amplitudes and remains only in the characteristic time-scales of the variations of the perturbed orbital elements. In practice, however, it is easy to see, that tight CHTs are the best candidates for observations of both types of perturbations. Therefore, in what follows, we discuss the main characteristics of these types of perturbations, and their relevance for the observations and the dynamical analyses of CHTs.

Before such an observation-oriented discussion, a very important caveat should be noted. The third-body perturbations in a triple system depend naturally, and exclusively, on the positions of the bodies relative to each other. Hence, these perturbations are generally described in a coordinate system which is anchored to the invariable plane of the triple (multiple) system. We will refer to this as the `dynamical system'. On the other hand, from an observational point of view, a natural coordinate system is that in which the basic plane is the tangential plane of the sky at the projected position of the target (i.e. the plane whose normal is the line of sight to the target). We will refer this latter as the `observational system'.  While those orbital elements that describe the size and shape of the (Keplerian) orbit, and the location of the body on that orbit (i.e., semi-major axis, $a$; eccentricity, $e$; and time of periastron passage, $\tau$) are independent of the coordinate systems, the angular orbital elements (inclination, $i$, argument of periastron, $\omega$, and longitude of the node $\Omega$) are different between the two coordinate systems. As an important consequence, those angular orbital elements of which the time-dependent variations have been studied, and described analytically (or, nowadays, often numerically), in perturbation studies cited above (and to be cited below) are not identical to the ones that are (or, can be) observed. For example, argument of periastron whose variation (i.e., apsidal motion) can be detected and measured through the shift of the secondary eclipses relative to the primary ones, is naturally the observational one ($\omega$), and should not to be confused with its dynamical counterpart ($\omega^\mathrm{dyn}$). The connection between the two apsidal advance rates is far from linear, and the observed apsidal advance rate (i.e. $\dot\omega$) does not depend exclusively on the dynamical apsidal advance rate ($\dot\omega^\mathrm{dyn}$) itself, but also depends  on the rate of nodal regression ($\dot\Omega^\mathrm{dyn}$) and, moreover, on the dynamical inclination ($i^\mathrm{dyn}$) as well (see Borkovits et al.~\cite{borkovitsetal07,borkovitsetal15}). In such a way, for example, in the most extreme cases, it may even happen that while in the dynamical system the argument of periastron rotates in a prograde direction (i.e. $\dot\omega^\mathrm{dyn}>0$), at the same time the observer detects a retrograde apsidal motion ($\dot\omega<0$). Similarly, the orbital plane precession in a non-coplanar triple manifests itself in an (almost) constant regression of the common nodal line ($\dot\Omega^\mathrm{dyn}<0$) of the two orbital planes in the invariable plane, while the inclinations of the orbital planes to the invariable plane ($i^\mathrm{dyn}$), may remain almost constant. Oppositely, in the observational system this effect may manifest itself in very large inclination variations of the EB, as will be discussed later.

\subsection{Long-period perturbations}
\label{sec:longperiod}

These perturbations were investigated for the first time analytically by S\"oderhjelm \cite{soderhjelm975,soderhjelm982,soderhjelm984}. His works are especially relevant here, as his main inspiration was to investigate the dynamics of two emblematic ECHTs, Algol and $\lambda$ Tauri. Besides formulating the long-period (and, also the apse-node) perturbations in terms of the usual Delaunay-type variables, he discussed some observational consequences for ECHTs. In regard to the long-period perturbations, perhaps his most important note is that ``true perturbations in the mean longitude enter directly in the times of minima. The apse-node perturbations are seldom important, but the long-period variation may be sensible.''\footnote{As we will discuss shortly below, for the tightest ECHTs, the apse-node perturbations also significantly influence the eclipse times.} 

The first order (quadrupole) analytical formulation of the effects of the long-period perturbations for the timing variations of circular inner EBs of ECHT-s was given by Mayer \cite{mayer990}. Later, in a series of papers Borkovits et al. \cite{borkovitsetal03,borkovitsetal11,borkovitsetal15} derived second order (octupole) formulae valid for ECHTs with arbitrarily eccentric inner and outer orbits. The significance of these perturbed timing variations in the regard of the detection and detailed characterization of ECHTs will be discussed below in Sect.~\ref{sec:ETVs}.

Besides the long-period perturbations of the timing variations of ECHT-s, there is another long-period perturbation effect that can be directly observed, at least by means of ultraprecise space-borne photometric observations. This effect manifests itself as periodic jumps in the eclipse depth variations of a few inclined ECHT-s, observed with the \textit{Kepler} and \textit{TESS} missions. As one can see in Fig.~\ref{fig:EDVs}, instead of smooth, linear variations, the eclipse depths in some EBs show staircase-like pattern where the jumps occur around the periastron passages of the third body. This is exactly due to the long-period perturbations in the observable inclination of the inner EB, which is superposed on the effect of the smoother, apse-node timescale orbital plane precession.

\subsection{Apse-node type perturbations}
\label{sec:apsenode}

From both the dynamical and evolutionary points of views, the most interesting and, therefore, more intensively studied perturbations are the apse-node timescale ones. The most well-known manifestation of such perturbations is the effect that have recently been referred to as the `von Zeipel-Lidov-Kozai effects' (hereafter ZLK) or `ZLK oscillations'. These phenomena were studied and described first in the pioneering works of von Zeipel~\cite{vonzeipel910}, Lidov~\cite{lidov962}, and Kozai~\cite{kozai962}. These authors originally investigated the motions of small bodies in triple systems (e.g. perturbations by Jupiter for comets and asteroids orbiting the Sun -- von Zeipel; Kozai; or perturbations by the Moon for the Earth-orbiting artificial satellites -- Lidov). In the context of hierarchical triple (and multiple stellar systems), ZLK phenomena have only been known and studied widely over the last $\sim25$\,years. The fact that these phenomena had remained nearly unknown to the double and multiple star community for more than 30 years may be a bit surprising. This is especially so in the light of the fact that in their first dedicated works about the stellar three-body problem, both Harrington~\cite{harrington968} and S\"oderhjelm~\cite{soderhjelm982} discuss in detail the analytic solution of Kozai's original formulae in regard to hierarchical triple star dynamics.  

According to these discussions, there are two domains of the mutual inclination angle of a hierarchical triple system limited by the value of $\cos^2i_\mathrm{mut}\approx\frac{3}{5}$.  In these two domains there is a substantial difference in the nature of the cyclic eccentricity variations of the inner binary\footnote{An exact equivalence occurs in the case of the asymptotic solution, i.e., when the orbital angular momentum is stored exclusively in a circular outer orbit}. For systems with low mutual inclinations (i.e. $i_\mathrm{mut}\lesssim39.23^\circ$ or, $i_\mathrm{mut}\gtrsim140.77^\circ$) there may occur only small-amplitude secular variations in the inner eccentricity (and the mutual inclination, as well). Furthermore, in the case of an initially circular inner orbit, it remains circular at all later times (of course, only as far as the approximation used remains valid).  In the high mutual inclination case, however, depending on the initial conditions, (i.e., the value of the inner binary eccentricity, mutual inclination and dynamical argument of pericenter at a given instant), the inner eccentricity may vary between zero and nearly unity, while the apsidal line may exhibit either circulation or libration.\footnote{Note, however, one can find specific solutions even for nearly perpendicular configurations, where the inner eccentricity remains (essentially) constant and the major axis of the orbit freezes into a specific direction.} 

Taking into consideration the octupole terms of the perturbation function and, allowing the third body to have any arbitrary eccentricity (e.g., Ford et al.~\cite{fordetal00}; Naoz et al.~\cite{naozetal13}), one can show that the problem is no longer analytically integrable, and the behaviour of the orbital elements may be more complex. For example, the relative orientations of the orbital planes may show flip-flop phenomena between prograde and retrograde, the rotation of the semi-major axis of the inner orbit may alternate between libration and circulation, and so forth. Detailed reviews of the ZLK phenomena can be found in Naoz~\cite{naoz16} and Ito \& Ohtsuka~\cite{itoohtsuka19}.

Though the large amplitude $e$-cycles of the ZLK phenomena are important from the point of view of the formation and also the later evolution of triple and multiple systems, from a direct observational point of view, especially in the case of (E)CHT-s, they have only some minor relevance. This is true because of the tidal forces that are present in short period close binaries and, therefore, in a large portion of the EBs. As was shown for the first time by S\"oderhjelm~\cite{soderhjelm984}, the tidally induced classical apsidal motion can decrease, or even completely negate, the third-body induced, large amplitude ZKL cycles for any mutual inclination. Note, this fact is crucial in regard to the survival of the triple system Algol in its present configuration. At the present time, this CHT is the only known system where the outer orbital plane is almost perpendicular to the plane of the close orbit of the semi-detached EB ($i_\mathrm{mut}=86\pm5^\circ$; Zavala et al.~\cite{zavalaetal10}). In the absence of the tidal effects, the inner orbit of Algol would be subjected to extremely large amplitude eccentricity variations which would lead to the merger of the two binary-star components within some thousand years.

Detailed investigations of the combined effects of third-body perturbations, i.e., ZKL cycles, and mutual tidal interactions (both conservative and dissipative) in hierarchical triple stellar systems have led Kiseleva et al.~\cite{kiselevaetal998} to propose the theory of `Kozai Cycles with Tidal Friction' (KCTF) as a promising explanation for the formation of the closest binary stars.\footnote{Note also the former pioneering work of Mazeh \& Shaham~\cite{mazehshaham79} which also addressed the connection between the third-body forced eccentricity cycles and tidal dissipation. This study, however, was restricted to only the low mutual inclination regime and, therefore was less general than that of the work of Kiseleva et al.~\cite{kiselevaetal998}.} According to this model, close binaries should have formed as inner binary subsystems of strongly inclined hierarchical triple systems. The initial separation of such pairs was substantially larger. Then, around the maximum eccentricity phase of the high mutual inclination generated, large amplitude $e$-cycles, the periastron distance of the inner binary temporarily decreased to such an extent that the mutual tidal friction along a section of the inner orbit became effective enough to shrink the orbit significantly. As a consequence, the tidal forces were able to act during increasingly longer sections of the inner orbit. Then, during one or more of the $e$-cycles, the tidal interactions became strong enough to decouple the inner binary from the ZKL-cycles, resulting in a highly eccentric inner binary with a moderate separation, and an inclined distant third stellar component. Finally, the tidal dissipation led to a further shrinkage and circularization of the inner binary orbit, producing the currently observed close circular binaries.

The implications of this theory were numerically tested by Fabrycky \& Tremaine~\cite{fabryckytremaine07} and Naoz \& Fabrycky~\cite{naozfabrycky14}, amongst others. They concluded that according to KCTF one can expect an enhancement of triple systems with inner periods of $P_\mathrm{in}=3-10$\,d and mutual inclinations of $i_\mathrm{mut}\sim40^\circ$ or $\sim140^\circ$. These predictions were later statistically investigated by Borkovits et al.~\cite{borkovitsetal16} in a sample of dynamically active ECHTs (see below, in Sect.~\ref{sec:ETVs}).

Later, however, Moe \& Kratter~\cite{moekratter18} and Tokovinin \& Moe~\cite{tokovininmoe20} have shown that KCTF cannot explain the observed fraction of close binary stars. Moreover, recent theories (see again, the review of Tokovinin~\cite{tokovinin21}) suggest that the majority of the most compact systems and, therefore, ECHTs should have formed in originally near planar stellar configurations. Hence, one cannot expect large amplitude $e$-cycles in CHTs, and this statement is in accord with our observational experience (see below).

The absence of large amplitude $e$-cycles, however, does not mean that apse-node timescale perturbations would be less important or observationally undetectable in ECHTs. To the contrary, in the tightest eccentric ECHTs the third-body forced apsidal motion in the inner binary may be extremely rapid. For example, in the case of KOI-126 Yenawine at al.~\cite{yenawineetal21} found an astonishingly rapid timescale of $P_\mathrm{apse}=1.74$\,yr for the apsidal motion. Moreover, two other ECHTs with apsidal motion periods shorter than 20 years (KIC~5771589; $P_\mathrm{apse}\approx6.5$\,yr; and KIC~7955301; $P_\mathrm{apse}\approx15$\,yr) were also reported in the \textit{Kepler} sample (Borkovits et al.~\cite{borkovitsetal15}). 

The contribution of forced negative dynamical apsidal motion from an unseen third companion was proposed as an explanation of the unexpectedly low apsidal motion rate observed in the eccentric EBs DI~Her and AS~Cam (Khaliullin et al.~\cite{khaliullinetal991}; Khodykin \& Vedeneyev~\cite{khodykinvedeneyev997}; Khodykin et al.~\cite{khodykinetal04}). Note, however, that later Albrecht et al.~\cite{albrechtetal09} found that the spin axes of both components of DI~Her are highly inclined to the normal of the orbital plane (similar to planet Uranus) which can explain the slow apsidal motion by itself through Shakura's model \cite{shakura985}. Pavlovski et al.~\cite{pavlovskietal11} arrived at a similar conclusion in the case of AS~Cam. Despite these findings, it remains a realistic future possibility to find third components in eccentric EBs through their effects on the apsidal motion of the EB (see also, Borkovits et al.~\cite{borkovitsetal19a}).

An immediate consequence of the non-alignment of the inner and outer orbital planes in triple systems, or the non-alignment of any two of the orbital planes of different subsystems in multiple systems, is the precession of the orbital planes. This effect in the dynamical frame of reference primarily manifests itself in nodal regression (at least for prograde configurations, i.e., when $i_\mathrm{mut}\leq90^\circ$). In the observational reference frame, however, its main consequence is the variation of the observed inclinations of both orbits, and therefore, in the case of an EB, it results in varying eclipse depths, and even the total disappearance of former eclipses or, oppositely, the occurrence of eclipses in formerly non-eclipsing binary stars. The characteristic time-scale of this effect for triples was given e.g. by S\"oderhjelm \cite{soderhjelm975} as

\begin{equation}
P_\mathrm{node}=\frac{4}{3}\frac{1+q_\mathrm{out}}{q_\mathrm{out}}\frac{P^2_\mathrm{out}}{P_\mathrm{in}}\left(1-e_\mathrm{out}^2\right)^{3/2}\left|\frac{C}{G_\mathrm{out}}\cos i_\mathrm{mut}\right|^{-1},
\label{eq:Prec}
\end{equation}
where $C$ represents the total orbital angular momentum of the triple, while $G_\mathrm{out}$ stands for the orbital angular momentum stored in the outer orbit. 

As one can see, the closer the orbital planes in a triple to the perpendicular configuration, the longer the nodal period. It is the reason why no eclipse depth variations have been detected in Algol, despite its relatively small ratio of $P_\mathrm{out}^2/P_\mathrm{in}\approx441$\,yr (S\"oderhjelm~\cite{soderhjelm980}). Orbital plane precession induced eclipse depth variations have been observed in some dozen CHTs, which will be discussed in Sect.~\ref{sec:depthvar}.

\vspace{0.3cm}

\section{Detection methods of CHTs}
\label{sec:detection_methods}

The 19 September 2021 update of the Multiple Star Catalog (MSC; Tokovinin~\cite{tokovinin18a})\footnote{\url{https://cdsarc.cds.unistra.fr/viz-bin/cat/J/ApJS/235/6} -- last accessed: 25 November 2021} contains $\approx201$ triple and 2+2 quadruple systems, or subsystems of higher order hierarchies with $P_\mathrm{out}\lesssim1000$\,days.\footnote{For comparison, we note that the former compilation of Fekel~\cite{fekel981} counted only 8 CHTs.} Out of these CHTs, 116 were found through detailed analyses of eclipse timing variations and/or, identifications of third-body eclipses of EBs observed by the \textit{Kepler} space telescope with its quasi-continuous and ultrahigh precision photometry over the four-year-long primary mission (Borucki et al.~\cite{boruckietal10}). Ten further CHTs were found in a similar manner via mining of the observational data of other space missions as CoRoT (Auvergne et al.~\cite{auvergneetal09}), \textit{K2} (Howell et al.~\cite{howelletal14}) and \textit{TESS} (Ricker et al.~\cite{rickeretal15}). To these, we can also add three additional triply eclipsing ECHTs found with \textit{TESS}, the analyses of which were published only recently (see below, in Sect.~\ref{sec:triply_eclipsers}) after the release of the last update of MSC. In all, we find that $\approx63$\% (129 of 204) of the known CHTs have been discovered in the last decade, through space photometry. Moreover, thanks to the continuous operation of the \textit{TESS} spacecraft, which looks at nearly all parts of the sky, further growth of this ratio is expected in the near future. The majority of the remaining 75 CHTs were found spectroscopically, through radial velocity (RV) analysis. 

In this section, we discuss the observational methods that are useful for the investigations of CHTs. We follow the historical order, i.e., we begin with spectroscopic observations or, more precisely, RV measurement and analysis (Sect.~\ref{sec:RVs}) which provided the most efficient way to discover and study such systems before the era of the exoplanet-hunter photometric space telescopes. Then, we continue with photometric methods, such as timing analysis of EBs (Sect.~\ref{sec:ETVs}), which may result in the detection of a close companion through the classical lightravel-time effect (LTTE), but also through the recently developed method of the identification and analysis of third-body perturbations. The latter are so-called `dynamical effects' (DE) in the timing variations which become significant in the tightest CHTs. In these systems the probability of third-body eclipses, though low, is not negligible. Therefore, it is not surprising that such phenomena have been observed in more than a dozen of the tight CHTs.  These will be discussed in Sect~.\ref{sec:triply_eclipsers}. Moreover, again, in the tightest CHTs the timescales of the longest apse-node timescale perturbations are short enough for the comfortable detection of their effects over human timescales. The most spectacular of these phenomena is the eclipse-depth variations observed in dozens of EBs, as will also be discussed in Sect.~\ref{sec:depthvar}. Finally, even though astrometry, the third large branch of classic astronomy, has only a lower impact in regard to CHTs, for the sake of completeness we review briefly astrometric measurements (mostly connected to interferometric techniques) of CHTs in Sect.~\ref{sec:astrometry}.

\subsection{Radial velocity measurements}
\label{sec:RVs}

As mentioned above, before the era of exoplanet hunter space telescopes, such as CoRoT (Auvergne et al.~\cite{auvergneetal09}), \textit{Kepler} (Borucki et al.~\cite{boruckietal10}; Howell et al.~\cite{howelletal14}) and its still operating successor, \textit{TESS} (Ricker et al.~\cite{rickeretal15}), most of the CHTs were discovered spectroscopically. This includes detecting either the periodically shifting absorption lines of a third star in a previously known spectroscopic binary or, even in the absence of signals from a third companion, the periodic variations in the RV curve of the binary itself.  The dominance of RV detection is in accord with the fact that, amongst the ground-based observation techniques, spectroscopic detection is the most effective for short period outer orbits (see, e.g., Mayer \cite{mayer990}, Tokovinin \cite{tokovinin14a}). The first spectroscopic detection of a tertiary component orbiting around an EB (or, more generally, around any binary system) was the discovery of the $P_\mathrm{out}=680$\,d period third star in the Algol system (Belopolski~\cite{belopolski906,belopolski911}, Curtiss~\cite{curtiss908}). Similarly, the currently shortest known outer period triple ($\lambda$~Tau; $P_\mathrm{out}=33.03$\,d; Schlesinger~\cite{schlesinger916}; Ebbighausen \& Struve \cite{ebbighausenstruve956}), and 2+2 type quadruple (VW~LMi; $P_\mathrm{out}=355$\,d; Pribulla et al.~\cite{pribullaetal06}) stellar systems were also discovered spectroscopically. Note also the currently tightest known 2+1+1 (or, planetary) type quadruple system, HIP~41431, of which the middle, $P_\mathrm{mid}=59$\,d period companion was discovered independently and simultaneously through spectroscopy (as a short period triple-lined, SB3, system) and photometry (eclipse timing variations during \textit{K2} observations). Moreover, the fourth component (having a period of $P_\mathrm{out}=1436$\,d) was also found spectroscopically (Borkovits et al.~\cite{borkovitsetal19b}).

In all of the above mentioned systems, additional stellar components of previously known EBs were found spectroscopically. However, a counter example is the case of HD~181068 (aka `Trinity'). Before the observations of the \textit{Kepler} spacecraft, the dominantly bright outer star of this system was categorized as a single-lined (SB1) spectroscopic binary (Guillout et al.~\cite{guilloutetal09}) with an orbital period of 45.5 days.  Its unique, triply eclipsing nature (see later, in Sect.~\ref{sec:triply_eclipsers}) was recognized only later, due to the unprecedently accurate photometric dataset of \textit{Kepler} (Derekas et al.~\cite{derekasetal11}). This is now understood to be one of the more compact ECHTs ($P_\mathrm{in}=0.9$\,d; $P_\mathrm{out}=45.5$\,d) found to date. The example of HD~181068 reveals some anti-selection bias against the spectroscopic, or even ground-based photometric, detection of such (E)CHTs in which the inner close binaries are the secondary components of previously known wider SB1 spectroscopic binaries.  It is simply because a much brighter outer component may dilute the light of the inner pair by such a large factor that its eclipsing nature may have remained unnoticed. For example, in the case of HD~181068 the red giant tertiary emits more than 99\% of the total flux of the triple system and, therefore, the amplitudes of the regular eclipses of the inner EB reduce to $\Delta{m}\approx0.003-0.004$\,mag which makes the chance of a serendipitous detection with ground-based instruments almost negligible.

Naturally, spectroscopically discovered CHTs do not necessarily contain an EB as their close, inner pair. However, there is a strong natural selection bias toward the spectroscopic discovery of ECHTs (and against non-eclipsing CHTs), because EBs, and especially short period ones, can be detected much more easily, and even serendipitously, compared with their non-eclipsing counterparts. And, moreover, short-period EBs are ideal and important targets for spectroscopic follow-up observations and, therefore, many of them were (and, are) observed and monitored systematically for RVs.

There are, however, some counter-examples as well. A systematic RV monitoring of hierarchical stellar systems carried out by Tokovinin~\cite{tokovinin16a,tokovinin16b,tokovinin18b,tokovinin18c,tokovinin19a,tokovinin19b, tokovinin20} has revealed a few CHT subsystems in some previously known, wider multiple systems.

Over the last decade, due not only to the above mentioned new space telescopes, but also to the large ground-based survey programs such as e.g. OGLE, ASAS-SN, and SuperWASP, there is a shift toward the photometric detection of ECHTs.  

It is important to note, however, that despite the recent dominance of the photometric detections of ECHTs, for the detailed, precise analysis of such systems, high-accuracy RV measurements continue to play a crucial role. They are the primary sources of model-independent, high precision stellar masses, which with the combinations of  high-level photometric measurements may provide other fundamental stellar parameters with high precisions, too. Recently, for example He{\l}miniak et al.~\cite{helminiaketal16,helminiaketal17,helminiaketal19} have been investigated systematically several single, double and triple lined EBs in the original field of the \textit{Kepler} spacecraft, including some ECHTs, too, that were discovered formerly through eclipse timing variations (see below, in Sect.~\ref{sec:ETVs}). Such a manner, they were able to determine, accurate masses, temperatures, metallicities, age and evolutionary status for some of the \textit{Kepler} ECHTs. 

\subsection{Eclipse Timing Variations}
\label{sec:ETVs}

A large fraction of third bodies orbiting around EBs were discovered through eclipse timing variation (ETV) studies.\footnote{Formerly, before the advent of extensive investigations of transiting exoplanets, any kind of timing variations in the binary star community were referred to almost exclusively as $O-C$ studies. This terminology stems from the use of {\em O}bserved minus {\em C}alculated diagrams in which one plots the observed times of eclipse minima (or, in the case of pulsating variables, maxima) minus the times calculated with the assumption of a constant period (see, e.g., Sterken~\cite{sterken05} for details). While terms like `$O-C$ diagram', `$O-C$ analysis' etc. continue to be in wide-spread use, in this review we apply the newest, and somewhat more accurate, terminology of `ETV'.} The two main sources of ETVs caused by an additional, distant, gravitationally bound body are the light-travel-time effect (LTTE), i.e. the classic R\o mer-delay and the third-body dynamical perturbations, which we will refer as the `dynamical effect' (DE). 

The LTTE produces a quasi-sinusoidal, strictly periodic signal with the period of the outer orbit ($P_\mathrm{out}$), and amplitude of
\begin{equation}
\mathcal{A}_\mathrm{LTTE}=\frac{a_\mathrm{AB}\sin i_\mathrm{out}}{c}\sqrt{1-e_\mathrm{out}^2\cos^2\omega_\mathrm{out}},
\label{eq:A_LTTE}
\end{equation} 
where $a_\mathrm{AB}$ denotes the semi-major axis of the EB (components A and B) around the center of mass of the triple system, $c$ is the speed of light, while the other symbols have their usual meanings. Analogous to its spectroscopic counterpart, by the use of Kepler's third law the mass function can be defined as

\begin{equation}
f(m_\mathrm{C})=\frac{m_\mathrm{C}^3\sin^3i_\mathrm{out}}{m_\mathrm{ABC}^2}=\frac{4\pi^2a_\mathrm{AB}^3\sin^3i_\mathrm{out}}{GP_\mathrm{out}^2}=\frac{c^3}{G}\frac{\mathcal{A}_\mathrm{LTTE}^3}{\left(1-e_\mathrm{out}^2\cos^2\omega_\mathrm{out}\right)^{3/2}}\left(\frac{2\pi}{P_\mathrm{out}}\right)^2.
\label{eq:f(mc)}
\end{equation}
Thus, the amplitude can be written as

\begin{equation}
\mathcal{A}_\mathrm{LTTE}\approx1.1\times10^{-4}f(m_\mathrm{C})^{1/3}P_\mathrm{out}^{2/3}\sqrt{1-e_\mathrm{out}^2\cos^2\omega_\mathrm{out}}\mbox{~[days]},
\label{eq:A_LTTE2}
\end{equation}
where the masses are in units of solar masses and the period in days. The amplitude of the LTTE can readily be converted into the amplitude of the systemic radial velocity variations of the EB as, 
\begin{equation}
K_\mathrm{AB}=\frac{\mathcal{A}_\mathrm{LTTE}}{P_\mathrm{out}}\frac{1\,883\,229}{\sqrt{\left(1-e_\mathrm{out}^2\right)\left(1-e_\mathrm{out}^2\cos^2\omega_\mathrm{out}\right)}}~\left[\mathrm{km\,s}^{-1}\right].
\label{eq:Kab}
\end{equation}

An LTTE solution of an ETV curve carries exactly the same information as can be inferred from the RV solution of a single-lined (SB1) spectroscopic binary. These include most of the orbital parameters of the outer orbit: $P_\mathrm{out}$, $e_\mathrm{out}$, $\omega_\mathrm{out}$, $\tau_\mathrm{out}$ (or, their equivalents) and the projected semi-major axis of the EB around the CM of the triple, $a_\mathrm{AB}\sin i_\mathrm{out}$, or, as an equivalent, the mass function $f(m_\mathrm{C})$. (Note, for the similar information content, LTTE dominated ETV and RV data can readily be converted into each other, as it was demonstrated by He{\l}miniak~\cite{helminiaketal16}.)

As was noted in the Introduction, Chandler~\cite{chandler892} was the first person who mentioned, and first formulated, the LTTE in the context of an explanation for the timing variations of Algol. Then, after the consecutive works of Hertzsprung \cite{hertzsprung922} and Woltjer~\cite{woltjer922}, Irwin \cite{irwin952,irwin959} gave the mathematical form of an ETV caused by LTTE in an easily applicable form that has been in widespread use since that time to the present.  Moreover, he also presented an easy to use graphical method to deduce the orbital parameters of the outer orbit from the ETV curves of the EBs.\footnote{In this regard we feel it necessity to make the following important cautionary note. In order to make his graphical solution simpler, Irwin \cite{irwin952} shifted the reference plane of the light-time orbit from the CM of the triple, i.e. the focal point of the outer ellipse to the geometric center of the orbit and thereby introduced an extra term of $a_\mathrm{AB}e_\mathrm{out}\sin\omega_\mathrm{out}\sin i_\mathrm{out}/c$ in his Eq. (2) (compare his Eqs.~[1] and [2]). This extra term has been also used in many of the recent papers dealing with LTTE. Besides the fact that there are no reasons to use this form in the era of numerical fitting procedures, an additional caveat is also warranted.  In particular, this step can be justified only insofar as the orbital elements of the light-time orbit remain constant. If the situation is in fact different (which may well occur in tight CHT-s, where the third-body perturbations can cause  variations of the orbital elements), this extra term would no longer remain constant and would lead to an error. Therefore, we recommend omitting the use of this additional $e_\mathrm{out}\sin\omega_\mathrm{out}$ term in all future studies. We made the same cautionary note in Borkovits et al.~\cite{borkovitsetal16}, but, unfortunately, we have found the same treatment in several newer papers as well.}  

\begin{figure}
\includegraphics[width=13cm]{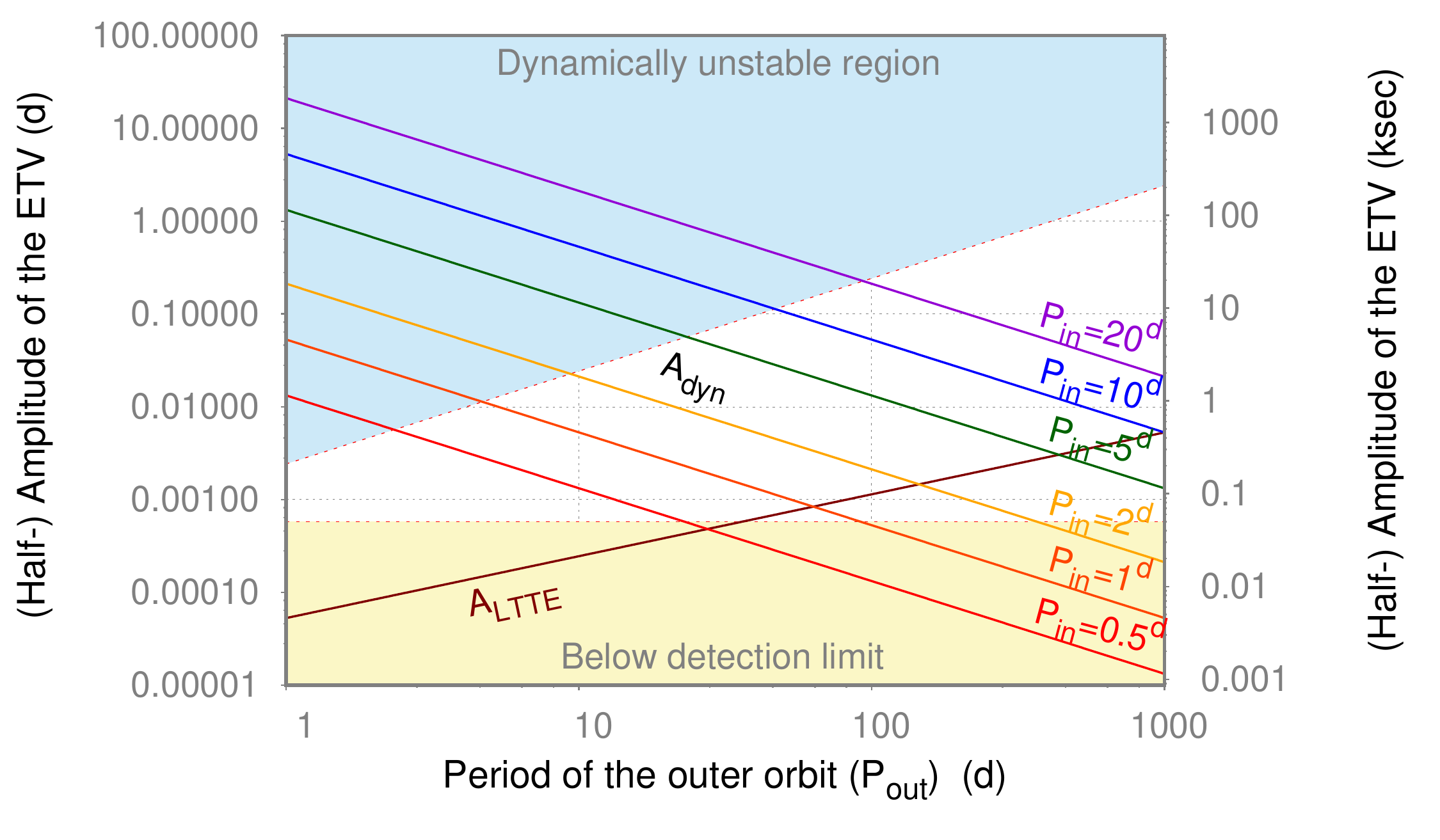}
\caption{The dependence of the LTTE and DE amplitudes ($\mathcal{A}_\mathrm{LTTE}$ and $\mathcal{A}_\mathrm{dyn}$, respectively) from the inner and outer orbital periods (adapted from Fig.~7 of Rappaport et al.~\cite{rappaportetal13}). The amplitudes were calculated with the use of formulae (\ref{eq:A_LTTE2}) and (\ref{eq:Adyn}). The masses of all the three components were set to $m_\mathrm{A}=m_\mathrm{B}=m_\mathrm{C}=1\,\mathrm{M}_\odot$, $e_\mathrm{out}=0.35$, $i_\mathrm{out}=60^\circ$, $\omega_\mathrm{out}=90^\circ$. The border of the dynamically unstable region was calculated according to the approximate formulae of Mardling \& Aarseth~\cite{mardlingaarseth01}. The limit of the certain detection was arbitrarily set to 50\,sec. \label{fig:A_LTTEDyn}}
\end{figure}

Some general criteria for accepting the plausibility of an LTTE model of ETVs have been given by Frieboes-Conde \& Herczeg~\cite{frieboescondeherczeg973}. Their criteria can be summarized as follows. (1) The shape of the ETV curve must follow the analytical form of an LTTE solution. (2) The amplitude and phase of the ETVs of the primary and secondary eclipses must be consistent with each other. (3) The estimated mass or lower limit to the mass of the third companion, derived from the amplitude of the hypothetical LTTE solution (see above), must be in accord with photometric measurements or limits on the third light contribution to the EB light curve. (4) Variation of the system RV should be in accord with the LTTE solution. While these criteria do not seem to be very restrictive, none of the 14 candidate EBs of \cite{frieboescondeherczeg973} fulfilled all of them. More than 70 years after the first mathematical description of the problem, there has been only one system, i.e. Algol itself, where the LTTE caused by the prior spectroscopically discovered third body (with period 680\,d in this case), was subsequently clearly confirmed via its ETV curve (Frieboes-Conde et al. \cite{frieboescondeetal970}). Even over the ensuing decades, the numbers of confirmed LTTE cases have grown very slowly. As discussed below, the main reason can be inferred from the mathematical form of the LTTE amplitude ($\mathcal{A}_\mathrm{LTTE}$, Eq.~[\ref{eq:A_LTTE2}]) and also, from Fig.~\ref{fig:A_LTTEDyn}. 

According to Eq.~(\ref{eq:A_LTTE2}), in the case of a hierarchical triple composed of three solar-mass stars, (i.e., $f(m_\mathrm{C})\lesssim0.11\,\mathrm{M}_\odot$) an LTTE results in an $\mathcal{A}_\mathrm{LTTE}\lesssim0.0027$\,d$\,\approx3.9$\,min (half-)amplitude ETV for $P_\mathrm{out}=1$\,yr, and $\mathcal{A}_\mathrm{LTTE}\lesssim0.0125$\,d$\approx18.1$\,min for $P_\mathrm{out}=10$\,yr. However, during the first $\sim70$\,years of the twentieth century most of the eclipse timing observations were done with visual brightness estimates having accuracies no better than a few hundredths of a day, and only a very limited number of photographic and photoelectric observations were available, and the latter were mainly concentrated on a few popular EBs (e.g., Algol itself).  It was, therefore, nearly hopeless to identify low amplitude light-travel-time orbits with periods shorter than a decade. Moreover, the ETVs of several EBs were found to show complex and sometimes erratic behaviour on different timescales (see, e.g. the $O-C$ atlas of Kreiner et al. \cite{kreineretal01}\footnote{http://www.as.up.krakow.pl/o-c/} for several examples).  These poorly understood variations may either act to hide short period LTTEs or, oppositely, mimic false LTTE signals, even on longer timescales (see, e.g., Hall~\cite{hall989}; Applegate~\cite{applegate992}; Lanza \& Rodon\`o~\cite{lanzarodono02}, for magnetic cycles; Kalimeris et al.~\cite{kalimerisetal02}; Tran et al.~\cite{tranetal13}; Balaji et al.~\cite{balajietal15}, for stellar spots and Borkovits et al.~\cite{borkovitsetal14}, for stellar oscillations).

Since the last decades of the past millennium, the advent of CCD detectors and other advances have led to the acquisition of much new and relatively accurate EB timing data.  These, in turn, have made possible the certain detection of LTTE in several dozens of EBs.  Most of these LTTE solutions reveal companions with orbital periods longer than a decade (see, e.g., Borkovits \& Heged\"us \cite{borkovitshegedus996}). Third stars were found in shorter period orbits only for a very limited number of EBs (see, e.g., Moffat et al.~\cite{moffatetal983}; Bartolini \& Zoffoli~\cite{bartolinizoffoli986}; Mayer~\cite{mayer990}; Chambliss~\cite{chambliss992}). In fact, before the observations made with the \textit{Kepler} spacecraft, IU~Aur was the only ECHT where a third-star companion with a period shorter than one year was discovered through LTTE (Mayer~\cite{mayer971,mayer983}). All the other tertiaries with periods less than one year had been discovered spectroscopically, as was discussed above, in Sect.~\ref{sec:RVs}. Note, however, after the conclusive spectroscopic detection of a third body, subsequent careful analyses of the ETVs have led to the identification of the LTTE in some of the most compact ECHTs, like DM~Per (Van Hamme \& Wilson~\cite{vanhammewilson07}) and VW~LMi (Pribulla et al.~\cite{pribullaetal08}). 

\begin{figure}
\includegraphics[width=4.5 cm]{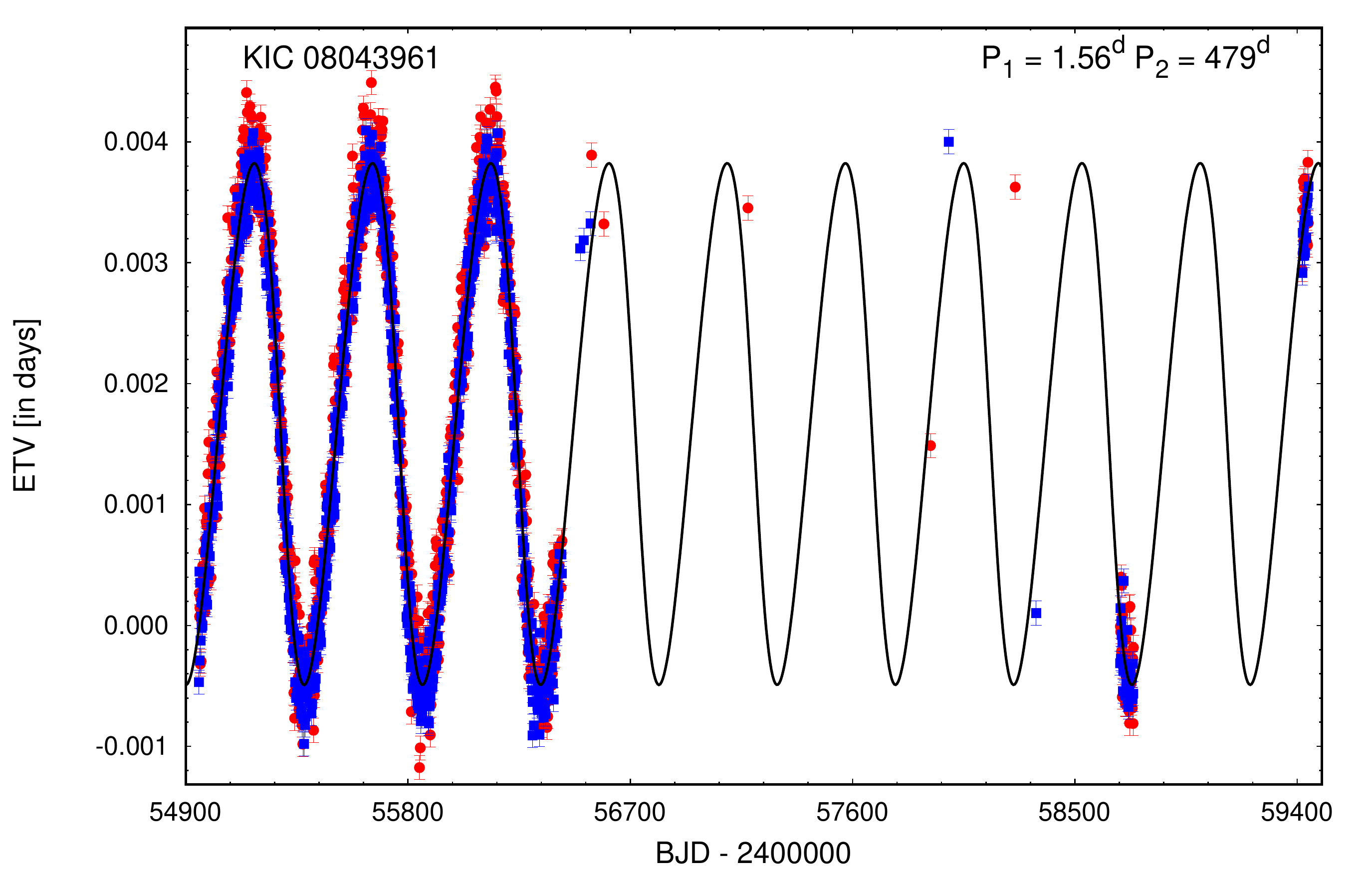}\includegraphics[width=4.5 cm]{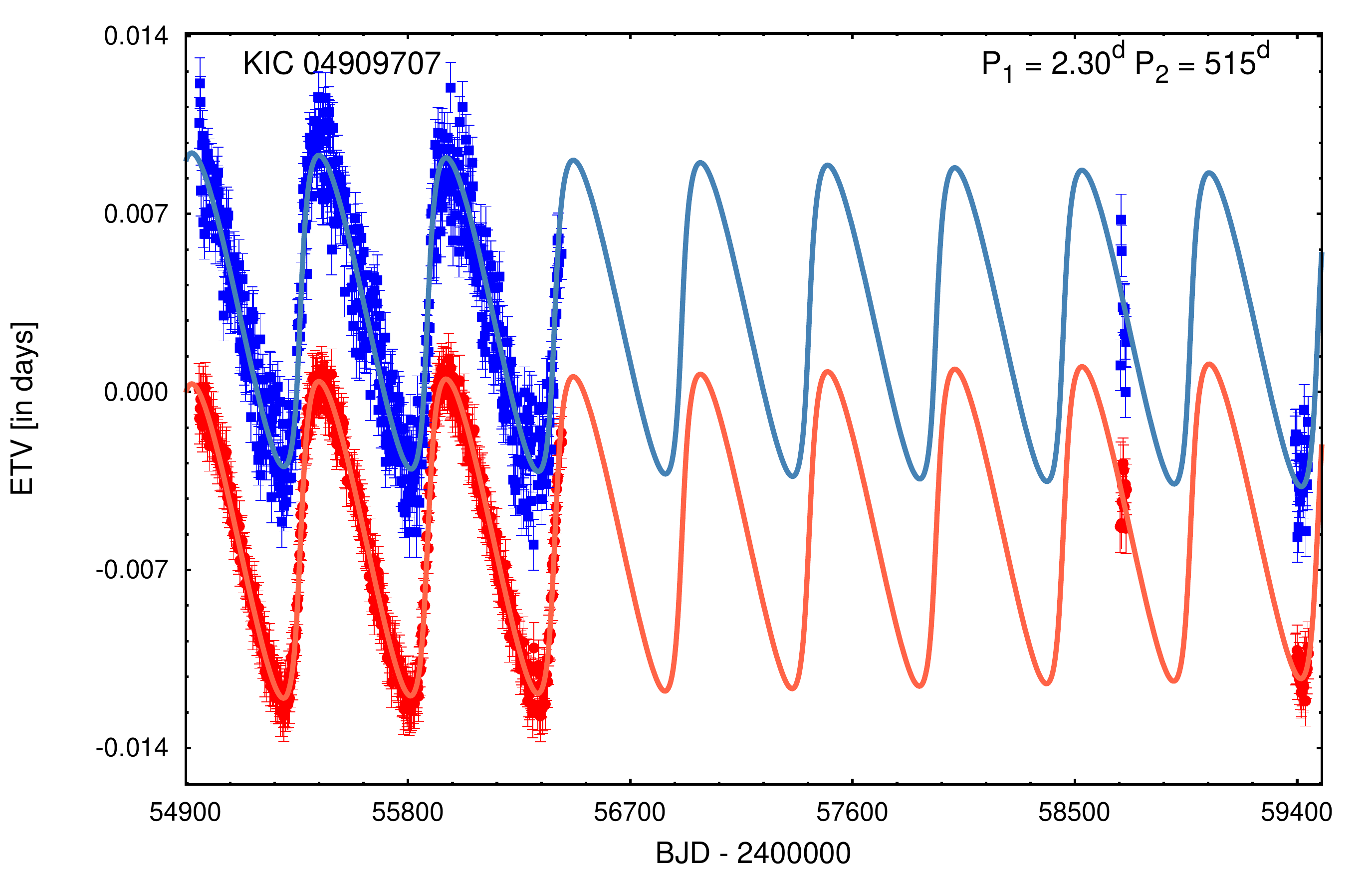}\includegraphics[width=4.5 cm]{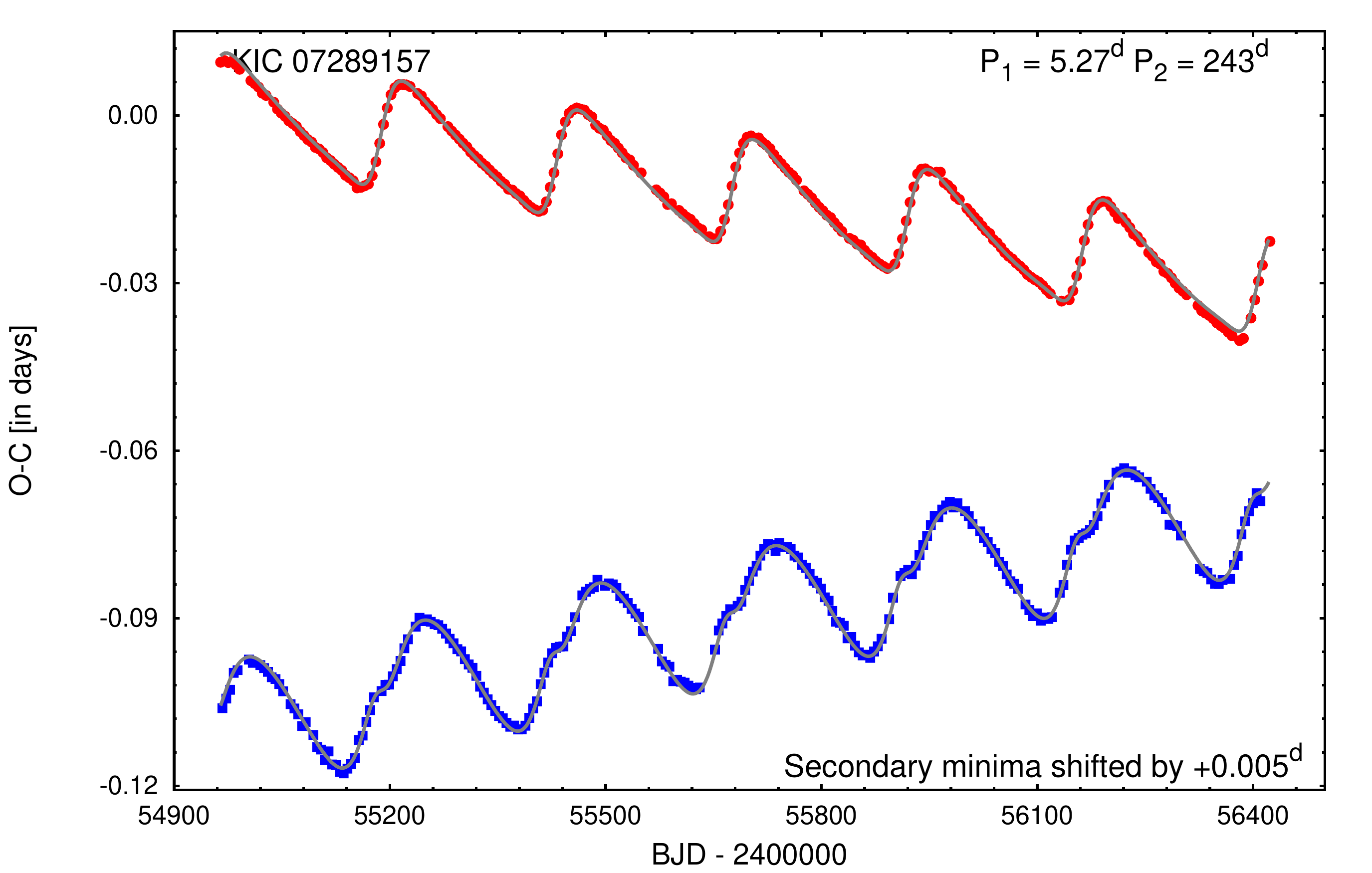}
\includegraphics[width=4.5 cm]{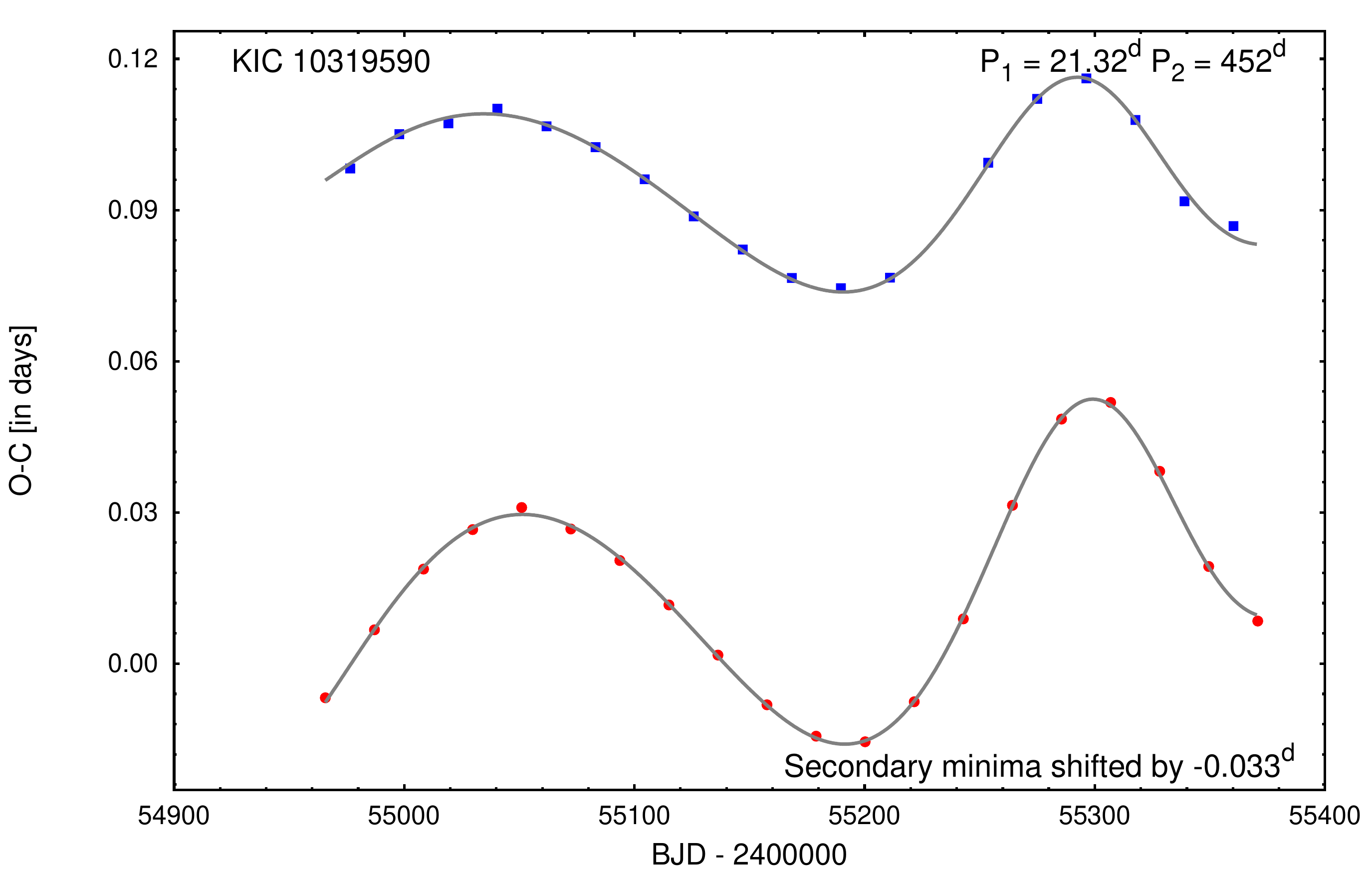}\includegraphics[width=4.5 cm]{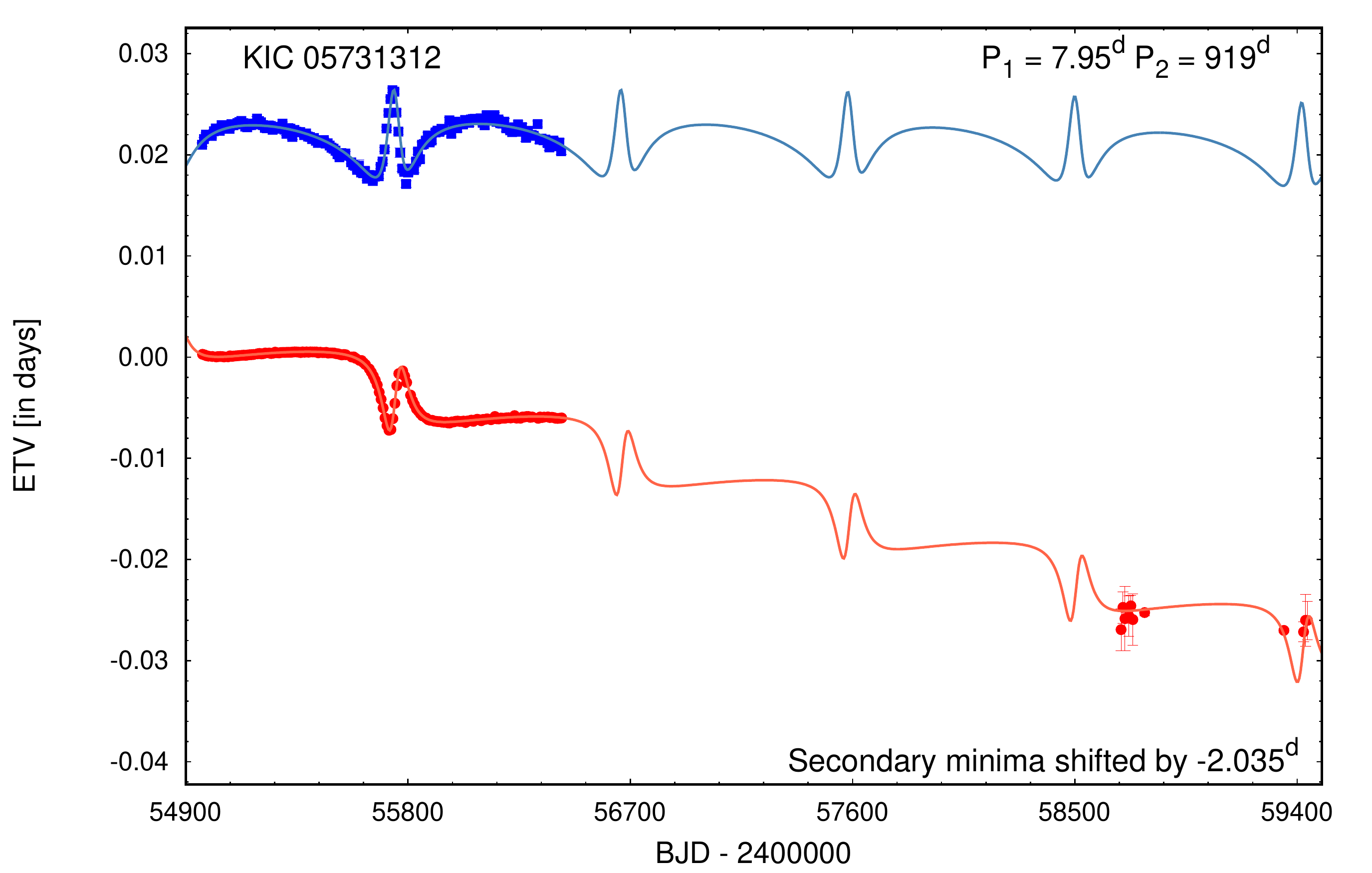}\includegraphics[width=4.5 cm]{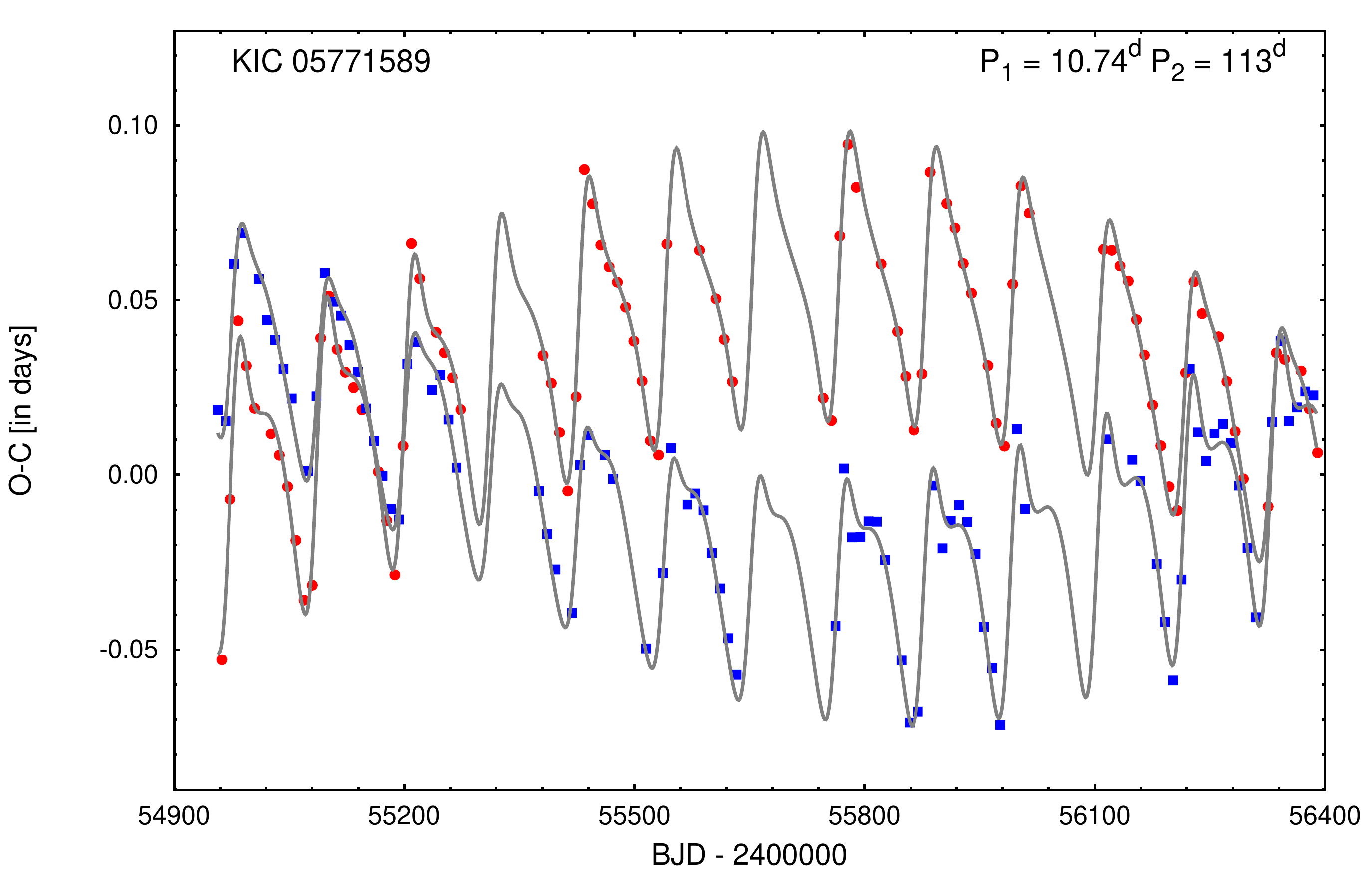}
\caption{Examples of ETVs amongst ECHTs in the original \textit{Kepler}-field. Primary and secondary eclipses are denoted by red circles and blue boxes, respectively. The continuous curves represent the analytic ETV solutions according to the model described in Borkovits et al.~\cite{borkovitsetal15}. The ETV curves from left to right, and then up to down represent the following ECHTs: KIC~8043961 ($P_\mathrm{in}=1.56$\,d; $P_\mathrm{out}=479$\,d): LTTE dominated ETV. The EB's orbit is circular (hence, there is no shift in the secondary minima), the outer orbit is moderately eccentric. (The few solo points in between the four-year-long \textit{Kepler} (to the left) and three sectors of \textit{TESS} data are (yet) unpublished minima points from the follow up photometry of Baja Observatory, Hungary.) -- KIC~4909707 (2.30;515): the LTTE and DE amplitudes are similar. The shift of the secondary eclipses indicates an eccentric inner orbit. (No apsidal motion is detectable.) The strong asymmetry of the ETV is a sign of a highly eccentric outer orbit. -- KIC~7289157 (5.27; 243): At first glance, this ETV mimics an LTTE, but it is clearly dominated by DE ($\mathcal{A}_\mathrm{DE}/\mathcal{A}_\mathrm{LTTE}\approx4.5$). Upon closer inspection one can see that the shape of the secondary curve differs significantly from the primary one, clearly violating the second criterion of Frieboes-Conde \& Herczeg~\cite{frieboescondeherczeg973}. The converging primary and secondary curves imply rapid apsidal motion. The eclipse depth of this triply eclipsing system decreased continually over the \textit{Kepler} lifetime and the system no longer exhibits eclipses during the \textit{TESS} era. (Hence, no ETV data can be extracted from the \textit{TESS} observations.) -- KIC~10319590 (21.32; 452): The ETV curves cover only one outer period, since the eclipses suddenly disappeared. The different amplitudes of the primary and secondary ETV curves, as well as their characteristic double-bumped nature, the latter of which is a common feature of moderately inclined ($i_\mathrm{mut}\approx40^\circ$) systems, are clearly visible. The LTTE contribution is practically negligible compared to the DE ($\mathcal{A}_\mathrm{DE}/\mathcal{A}_\mathrm{LTTE}\approx10$). -- KIC~5731312 (7.95; 919): Due to the higher inner and outer eccentricities, one of the two ETV bumps turns into a sharp spike (indicating the periastron passage of the outer body), while the other bump becomes much shallower. The shape of the two curves differs strongly. Moreover, the apsidal motion of the inner EB is clearly detectable. The eclipse depth is decreasing continuously. By the time of the \textit{TESS} measurements the secondary eclipses are completely missing. Primary eclipses will also disappear in the near future. -- KIC~5771589 (10.74;113): One of the tightest ECHT in the \textit{Kepler} sample. The extremely rapid apsidal motion ($P_\mathrm{apse}\approx6.5$\,yr) is quite apparent. This ECHT has also a similarly short nodal period ($P_\mathrm{node}\approx7.5$\,yr). \label{fig:ETVs}}
\end{figure}

The new epoch of precision space-borne photometry, however, revolutionized the discovery of very short outer period ECHTs via eclipse timing analyses. The \textit{Kepler} EB observations were utilized right from their beginnings to search for signals of third companions amongst ETV data. Early attempts were done by Gies et al.~\cite{giesetal12} who identified possible long-term ETVs in 14 of 41 EBs but did not find evidence of short-period companions ($P_\mathrm{out}<700$\,d). Later they improved their analysis and obtained certain LTTE solutions for seven EBs, however all but one outer period was found to be much longer than 1000\,days. A more comprehensive analysis was carried out by Rappaport et al.~\cite{rappaportetal13} who identified 39 ECHTs in the \textit{Kepler} sample. In addition to the discovery of a significant number of new ECHTs, the main importance of this paper, however, was their demonstration that in a significant number of \textit{Kepler} EBs, the previously generally neglected dynamical effects must also be taken into account (see Fig.~\ref{fig:ETVs}). (We will return to the dynamical effects, soon.)  Another detailed analysis was carried out by Conroy et al.~\cite{conroyetal14} who determined eclipse times for all the short-period EBs of the original \textit{Kepler} sample, and identified 236 systems for which the ETVs can be modeled with the LTTE. Though the majority of their 236 LTTE candidate systems were observed for less than one complete outer period (or, in other words, the derived outer periods were found to be longer than the $\sim4$\,yr-duration of the dataset), some triple candidates with outer periods $P_\mathrm{out}\lesssim1000$\,d were also found. Another study was carried out by Zasche et al.~\cite{zascheetal15} who investigated 10 EBs in the \textit{Kepler}-field searching also for LTTE. 


Parallel to the revolution of space-photometry, the era of large, quasi-continuous ground-based surveys was ushered in. These surveys featured long-term, highly homogenous, night-by-night observations, which made it possible to find, in addition to their main goals of discovering such things as exoplanets and gravitational microlensing, new triple stars amongst EBs (most of them previously unknown) including a number of ECHTs as well. 

With the combination of photometric data from the surveys of MACHO (Derekas et al.~\cite{derekasetal07}; Faccioli et al.~\cite{facciolietal07}), OGLE-II (Wyrzykowski et al.~\cite{wyrzykowskietal04}), OGLE-III (Pawlak et al.~\cite{pawlaketal13}), OGLE-IV (Pawlak et al.~\cite{pawlaketal16}) and their own new CCD observations, Zasche et al.~\cite{zascheetal16,zascheetal17} investigated, for the first time, the LTTE in bright EBs in the Large and Small Magellanic Clouds. They found 22 hierarchical triple or quadruple candidate systems, from which two have outer periods shorter than 3\,years.

Hajdu et al.~\cite{hajduetal19,hajduetal22} studied 425,193 EBs towards the galactic bulge monitored during the photometric survey of OGLE-IV (Soszy\'nski et al.~\cite{soszynskietal16}). Among the 80\,000 EBs that they selected for detailed investigation, they identified 992 potential hierarchical triple (or quadruple) candidates of which the ETVs most probably exhibit LTTE (or, in a few cases, a combination of LTTE and DE) out of which 258 are categorized as certain candidates, most of them ECHT-s (i.e., $P_\mathrm{out}\lesssim1000$\,d).

Other ECHTs were identified via LTTE within the frame of the Microlensing Observations in Astrophysics (MOA-II) project (Li et al.~\cite{lietal17,lietal18}).

Last, but not least, we also mention the results of Wolf et al.~\cite{wolfetal16} which do not connect to any of the above mentioned large surveys but, instead involve long-term, targeted monitoring of EBs by groups of serious amateur astronomers using appropriate techniques. They have identified third stellar components with ETV analyses of low-mass EBs, one of which (NSV~07453183) has an outer period as short as $P_\mathrm{out}=418$\,d.

In addition to LTTE, third-body perturbations in the two-body motion of a close eclipsing pair in a tight ECHT may also produce readily observable ETVs. These effects in general manifest themselves in the ETV curves on two substantially different timescales in accord with the categories of long-period and apse-node timescale perturbations (see Sect.~\ref{Sect:dynamics}) as their origins.

\begin{figure}
\includegraphics[width=13cm]{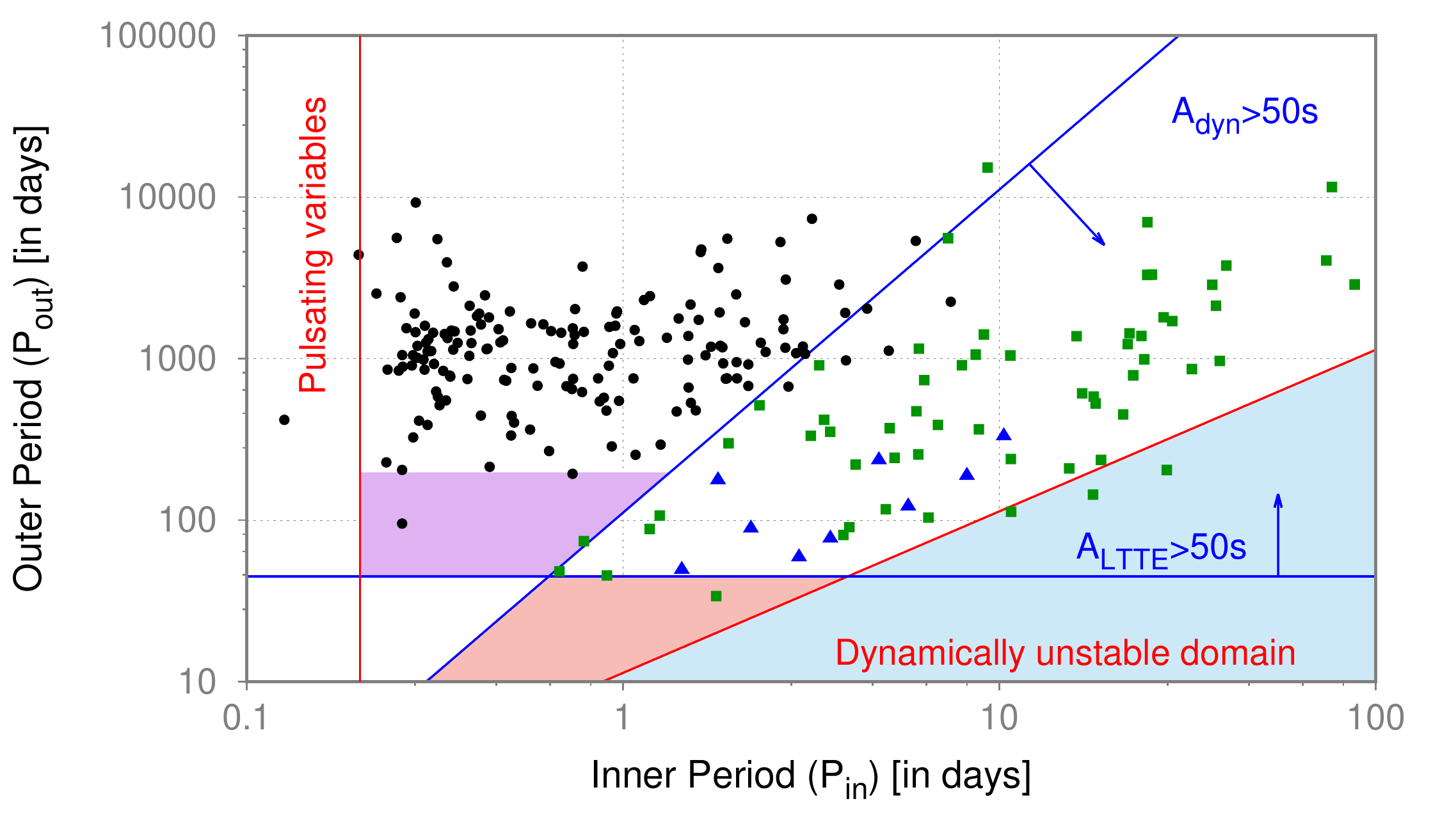}
\caption{Locations of CHTs detected with space telescopes \textit{Kepler} and \textit{TESS} in the $P_\mathrm{in}-P_\mathrm{out}$ plane (updated from Fig.~8 of Borkovits et al.~\citealt{borkovitsetal16}).  Black dots and green squares represent those 222 CHTs of the original \textit{Kepler}-field which were analysed in Borkovits et al.~\cite{borkovitsetal16}. For the black systems simple LTTE solutions were satisfactory, while the green ones the dynamical effects were also taken into account.  Blue triangles represent recently discovered \textit{K2} and \textit{TESS} systems with accurately known photodynamical solutions. The vertical red line at the left shows the  lower limit of the period of overcontact binaries. The horizontal and sloped blue lines are boundaries that roughly separate detectable ETVs from the undetectable ones. The detection limits again, were set to 50\,sec. These amplitudes were calculated with the similar assumptions than that of in Table~\ref{fig:A_LTTEDyn}. The arrows indicate the direction of the increasing of the respective amplitudes. The shaded regions from left to right represents $(i)$ the W~UMa desert, i.e. the (almost) empty domain where a tight third companion of a short-period (mostly W UMa-type overcontact) EB would be certainly detectable through its LTTE; $(ii)$ the purely dynamical region, i.~e., where the dynamical effect should be detectable, while the R\o mer-delay not and; $(iii)$ the dynamically unstable domain in the sense of the Mardling \& Aarseth~\cite{mardlingaarseth01} formula. \label{fig:P1vsP2}}
\end{figure}   

Long-period third-body perturbations cause ETVs that, in most cases, are similar to the LTTE in that they are periodic with the outer orbital period ($P_\mathrm{out}$). However, there may also occur some special orbital configurations that lead to ETVs with a period of $P_\mathrm{out}/2$. In contrast to the LTTE, the analytic form of DEs cannot be described with a pure sinusoidal term and, therefore, its amplitude also cannot be given in a simple closed form. According to the investigations of Borkovits et al.~\cite{borkovitsetal11,borkovitsetal15} the DE generated ETV curves have complicated dependencies on both the inner and outer eccentricities ($e_\mathrm{in,out}$), the mutual inclination of the two orbital planes ($i_\mathrm{mut}$) and, moreover, on both the observed and the dynamical arguments of periastron of the two orbits ($\omega_\mathrm{in,out}$ and $\omega_\mathrm{in,out}^\mathrm{dyn}$, respectively).  Note, these complex dependencies may result in strongly different amplitudes for the primary and secondary ETV curves in the same EB. Despite these complications, Borkovits et al.~\cite{borkovitsetal16} found that the amplitude of DE-caused ETV in many cases can be well approximated with the simple expression:
\begin{equation}
\mathcal{A}_\mathrm{dyn}\approx\frac{1}{2\pi}\frac{m_\mathrm{C}}{m_\mathrm{ABC}}\frac{P_\mathrm{in}^2}{P_\mathrm{out}}\left(1-e_\mathrm{out}^2\right)^{-3/2},
\label{eq:Adyn}
\end{equation}
or, for nearly coplanar (flat) ECHTs, with small inner eccentricities, according to Rappaport et al.~\cite{rappaportetal13} an even better approximation is
\begin{equation}
\mathcal{A}_\mathrm{dyn}^\mathrm{flat}\approx\frac{3}{8\pi}\frac{m_\mathrm{C}}{m_\mathrm{ABC}}\frac{P_\mathrm{in}^2}{P_\mathrm{out}}\left(1-e_\mathrm{out}^2\right)^{-3/2}e_\mathrm{out}.
\label{eq:Adynflat}
\end{equation}

Note, as the second form directly suggests, both Borkovits et al.~\cite{borkovitsetal03} and Agol et al.~\cite{agoletal05} have found that the DE disappears for coplanar systems (i.e. $i_\mathrm{mut}=0^\circ,~180^\circ$) where both the inner and the outer orbits are circular ($e_\mathrm{in}=e_\mathrm{out}=0$). Though later Borkovits et al.~\cite{borkovitsetal15} have shown that this statement is strictly valid only insofar as the octupole and higher order terms of the perturbing forces are neglected.  However, these higher order, non-vanishing DE contributions in the flat, doubly circular case will remain small, and turn out to be negligible in practice. This scenario is realized in at least three known triply eclipsing ECHTs (HD~181068, Borkovits et al.~\cite{borkovitsetal13}; TIC 278825952, Mitnyan et al.~\cite{mitnyanetal20}; and TIC 193993801, Borkovits et al.~\cite{borkovitsetal22}), where numerical integrations have justified that, despite the tightness of these systems, their observed ETVs are dominated exclusively by LTTE.

As one can see, in general the amplitude of the DE, besides its dependence on the tightness of the system ($P_\mathrm{in}/P_\mathrm{out}$), and on the outer mass ratio $m_\mathrm{C}/m_\mathrm{ABC}=q_\mathrm{out}/(1+q_\mathrm{out})$ (which terms are common to the amplitudes of the long-period variations of all the orbital elements of the EB), is also scaled  by the EB's period ($P_\mathrm{in}$). As a direct consequence, the longer the inner period, the larger and, therefore, better observable are the DE-caused timing variations (see, again, Fig.~\ref{fig:A_LTTEDyn}).  What follows from this is that for short-period EBs, which are ideal for ground-based photometric eclipse observations and, therefore, timing measurements, this effect always remains small (in almost all the known cases much smaller than LTTE). This explains why the DE was almost never considered, with only a very few exceptions before the era of \textit{Kepler}. 

Apart from some theoretical investigations (discussed in Sect.~\ref{sec:longperiod} above), the first attempt to incorporate DE into the interpretation of the observed ETVs of EBs (IU~Aur and FZ~CMa) was made by Mayer~\cite{mayer990} (see, also, \"Ozdemir et al.~\cite{ozdemiretal03}, for an updated analysis of IU~Aur.). However, even in these two ECHTs, the LTTE substantially dominates over the DE. 

The DE has become unavoidable in the ETV analysis of several, longer period EBs in the original \textit{Kepler}-field. Many of these EBs, even upon casual inspection, show ETV patterns which evidently cannot be explained with LTTE (see Fig.~\ref{fig:ETVs}), but can nicely be modeled with DE. The first ETV analysis of a DE dominated ECHT, KOI-928 was carried out by Steffen et al.~\cite{steffenetal11}. Then Rappaport et al.~\cite{rappaportetal13} surveyed the $Q0-Q13$ \textit{Kepler} quarter datasets for about 2100 EBs, and identified 39 short outer period ECHTs for which they presented combined LTTE+DE solutions. In these first works, however, as a simplification, the orbits of the inner EBs were considered to be circular. The eccentricity of the inner orbit, and also the additional, apse-node timescale DE-s (see Sect.~\ref{sec:apsenode}) were incorporated into the analysis, for the first time by Borkovits et al.~\cite{borkovitsetal15}, who investigated 26 eccentric EBs found with {\it Kepler}, all of them with DE dominated ETVs.  They showed that the DE solutions may be used to determine some important dynamical parameters of the systems, including the mutual inclination ($i_\mathrm{mut}$) which is a key-parameter in the dynamical evolutions of CHT-s (see Sect.~\ref{sec:apsenode}, above).  Moreover, they found that in an ideal case, the individual masses of the stars can also be estimated from the combined ETV analysis. 

Borkovits et al.~\cite{borkovitsetal16} then used the above LTTE plus DE generalized analyses (i.e., including eccentricities for both orbits, mutually inclined orbits, and apse-node timescale contributions) to study the complete sample of the Villanova \textit{Kepler} EB catalog\footnote{\url{http://keplerebs.villanova.edu/}} (Pr\v{s}a et al.~\cite{prsaetal11}; Slawson et al.~\cite{slawsonetal11}; Matijevi\v{c} et al.~\cite{matijevicetal12}; Conroy et al.~\cite{conroyetal14}). 
After filtering out false positives, they obtained certain, or probable, combined LTTE+DE solutions for 62 tight triple systems and, moreover, they found pure LTTE solutions for 160 additional EBs.  For the 62 triples with combined LTTE and DE solutions they were able to determine the mutual inclinations of the inner and outer orbital planes and found that the vast majority of these tight systems have low $i_\mathrm{mut}$, i.e., these are mostly flat systems.  This is in accord with the theories of the formation of CHTs through disc fragmentation (see, e.g., Tokovinin \& Moe~\cite{tokovininmoe20}).\footnote{A second peak in $i_\mathrm{mut}$ was also found, around $\sim40^\circ$, which coincides with the predictions of the numerical simulations of Fabrycky \& Tremaine~\cite{fabryckytremaine07} based on the KCTF mechanism (Sect.~\ref{sec:apsenode}).  However, the distribution of the inner periods does not follow the parallel prediction of the same model and, therefore, the mentioned agreement is most probably incidental.}  Amongst their 222 hierarchical triple or multiple systems (or, candidates), Borkovits et al.~\cite{borkovitsetal16} identified 104 ECHTs (i.e., following the strict definition of systems with $P_\mathrm{out}\lesssim1000\,$d). They found that the distribution of the outer periods is more or less flat over the regime of $200\lesssim P_\mathrm{out}\lesssim1600$, of which the lower boundary implies the existence of the `W~UMa desert', i.e., the absence of tight compact systems amongst the overcontact (mostly W~UMa-type) binaries (see Fig.~\ref{fig:P1vsP2}), as was previously noted by Conroy et al.~\cite{conroyetal14}.

A similar study was carried out by Hajdu et al.~\cite{hajduetal17} to investigate CHTs in the fields observed with the CoRoT satellite. Due to the much shorter observational datasets, their findings are less certain, however, they did identify six CHTs in the period regime of $82\leq P_\mathrm{out}\leq850$\,d. One of the six candidates, CoRoT~104079133 was also found to be a triply eclipsing triple star.

Recently, discoveries of the first CHTs via the ETV method in the \textit{TESS} sample were also reported.  TIC~167692429 ($P_\mathrm{in}=10.26$\,d; $P_\mathrm{out}=331$\,d) and TIC~220397947 ($P_\mathrm{in}=3.55$\,d; $P_\mathrm{out}=77$\,d) are located near to the southern continuous viewing zone (SCVZ) of \textit{TESS} spacecraft, and therefore they were observed almost continuously during Years 1 and 3 of the mission (Borkovits et al.~\cite{borkovitsetal20b}). Some other ECHTs with significant LTTE and/or DE in their ETVs were also discovered and investigated during the last two years, however, as these systems were originally found through the detection of extra eclipses in their light curves, we leave their discussion to Sect.~\ref{sec:triply_eclipsers}.

\subsection{Extra eclipses}
\label{sec:triply_eclipsers}

One of the most spectacular features that can be observed in a few EBs is the occurrence of `extra' or anomalous eclipses in their light curves. When these recur strictly periodically,  or nearly so, the most probable origin are the eclipses of a third (or more) stellar or substellar body.  These may be either a third star (or, circumbinary planet) which forms a hierarchical triple system with the inner EB or, another EB being either bound in a 2+2 quadruple configuration or unbound to the first EB. The two scenarios (third-body eclipses vs double EB) in most cases can be readily distinguished upon a first inspection. 

\subsubsection{Third-body eclipses}

\begin{figure}
\includegraphics[width=4.5 cm]{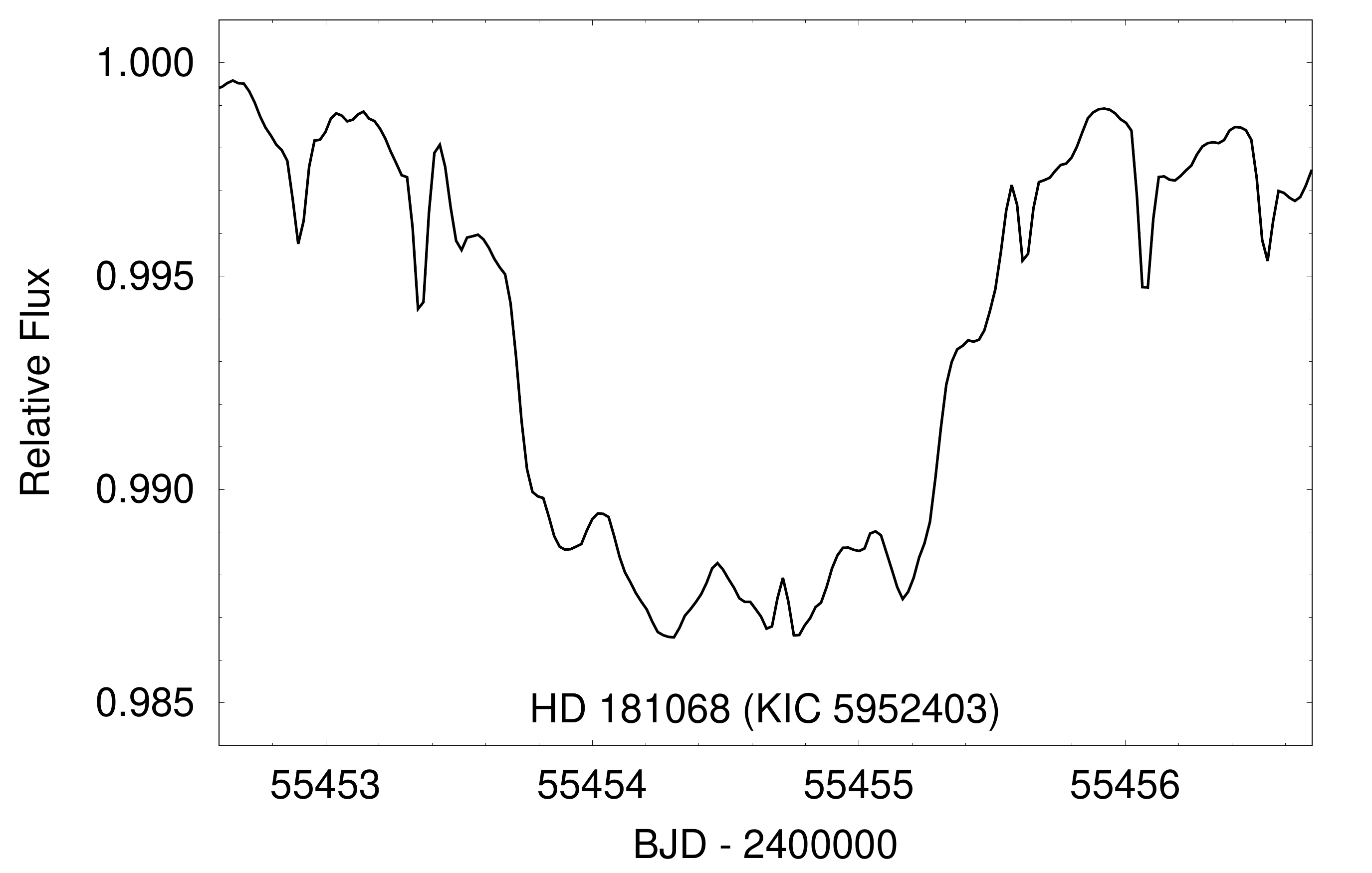}\includegraphics[width=4.5 cm]{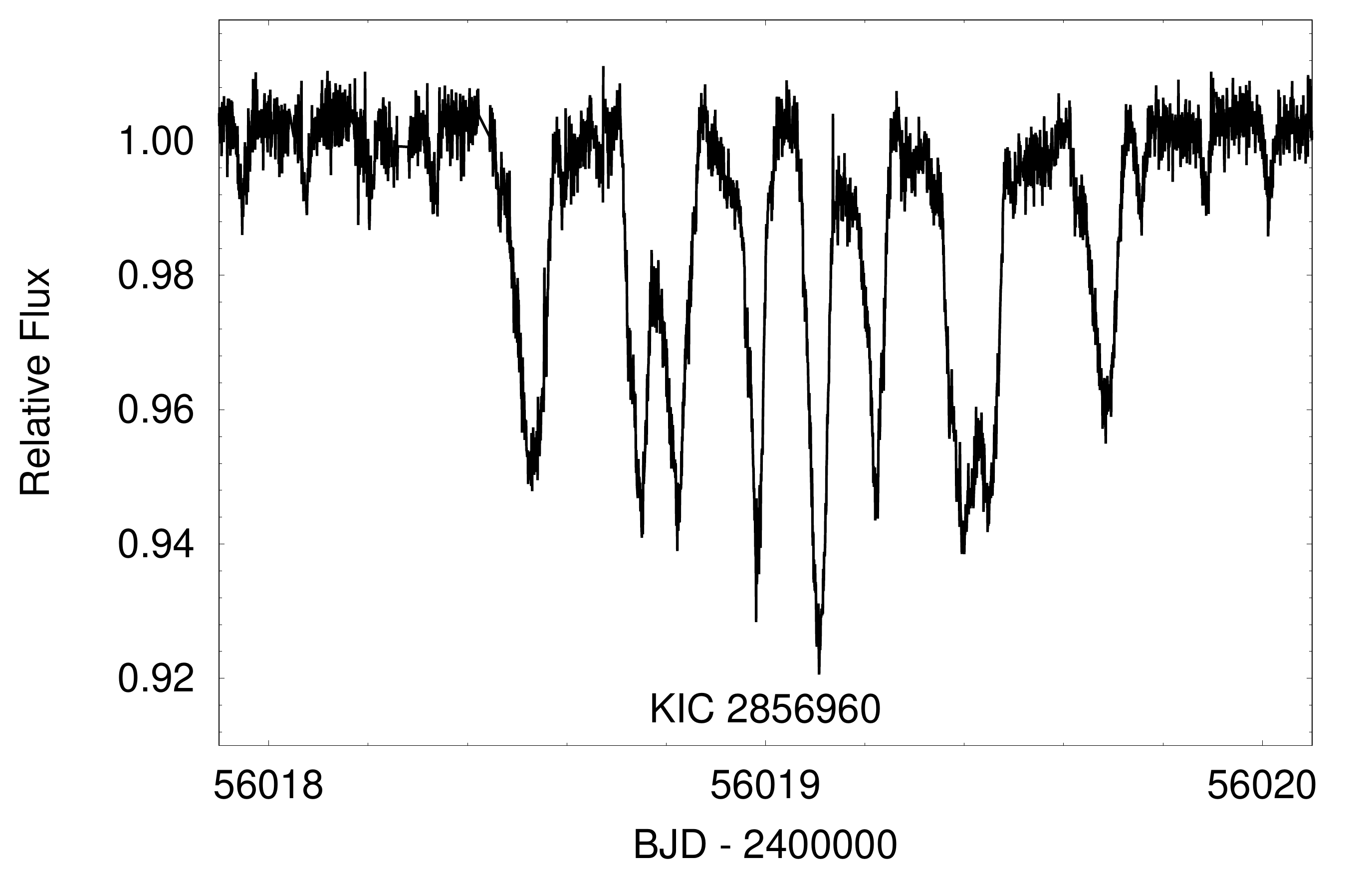}\includegraphics[width=4.5 cm]{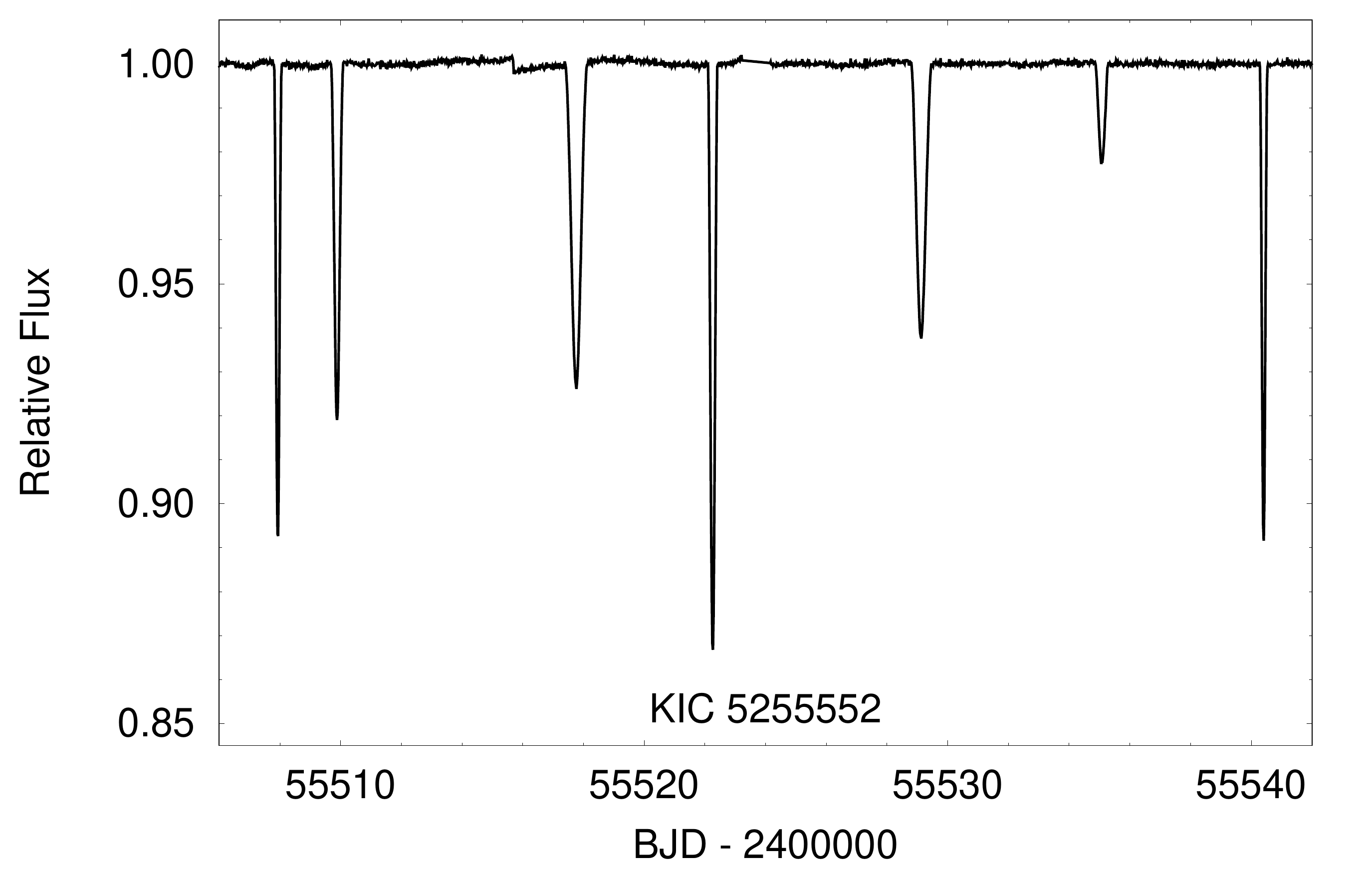}
\includegraphics[width=4.5 cm]{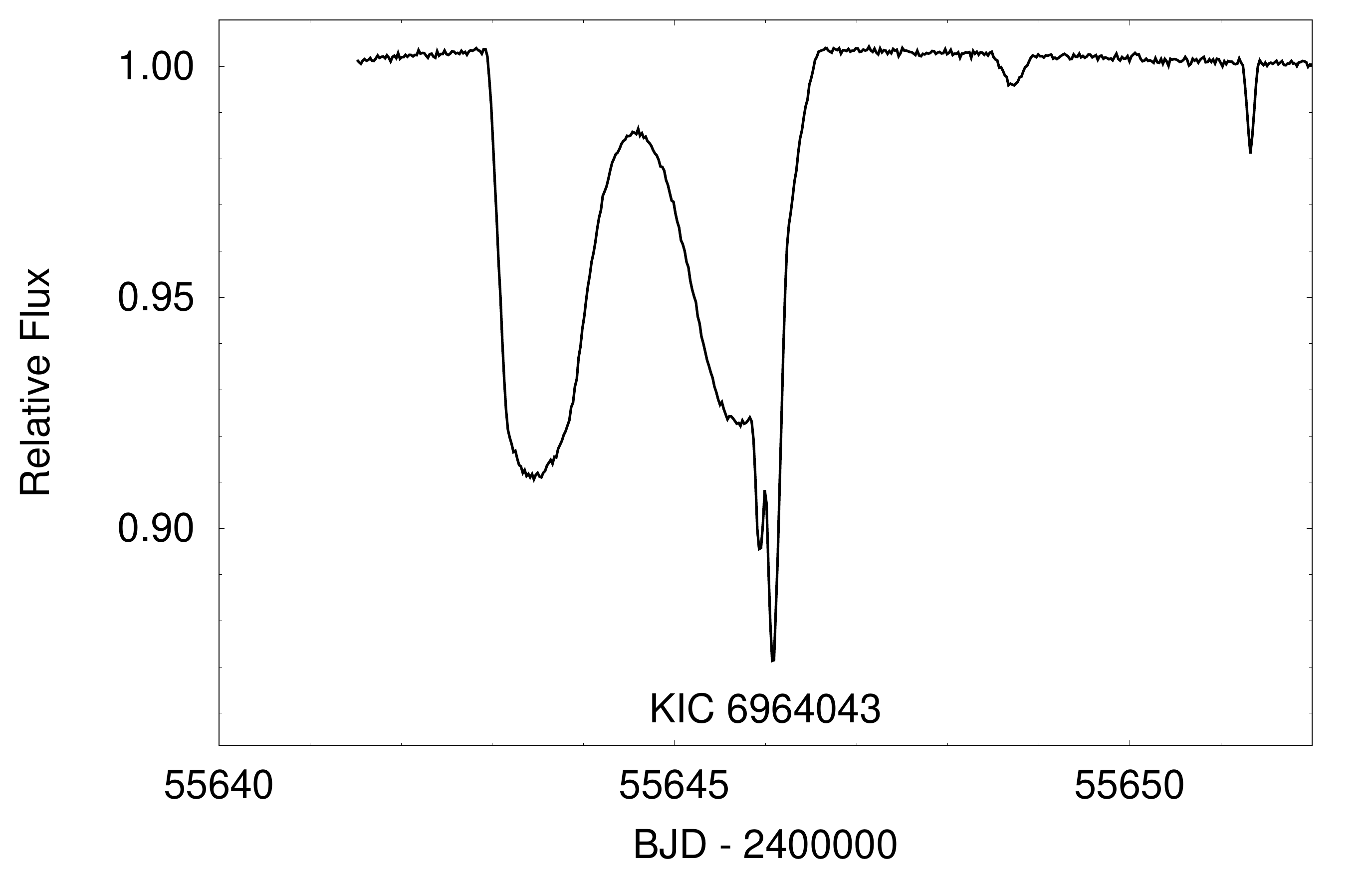}\includegraphics[width=4.5 cm]{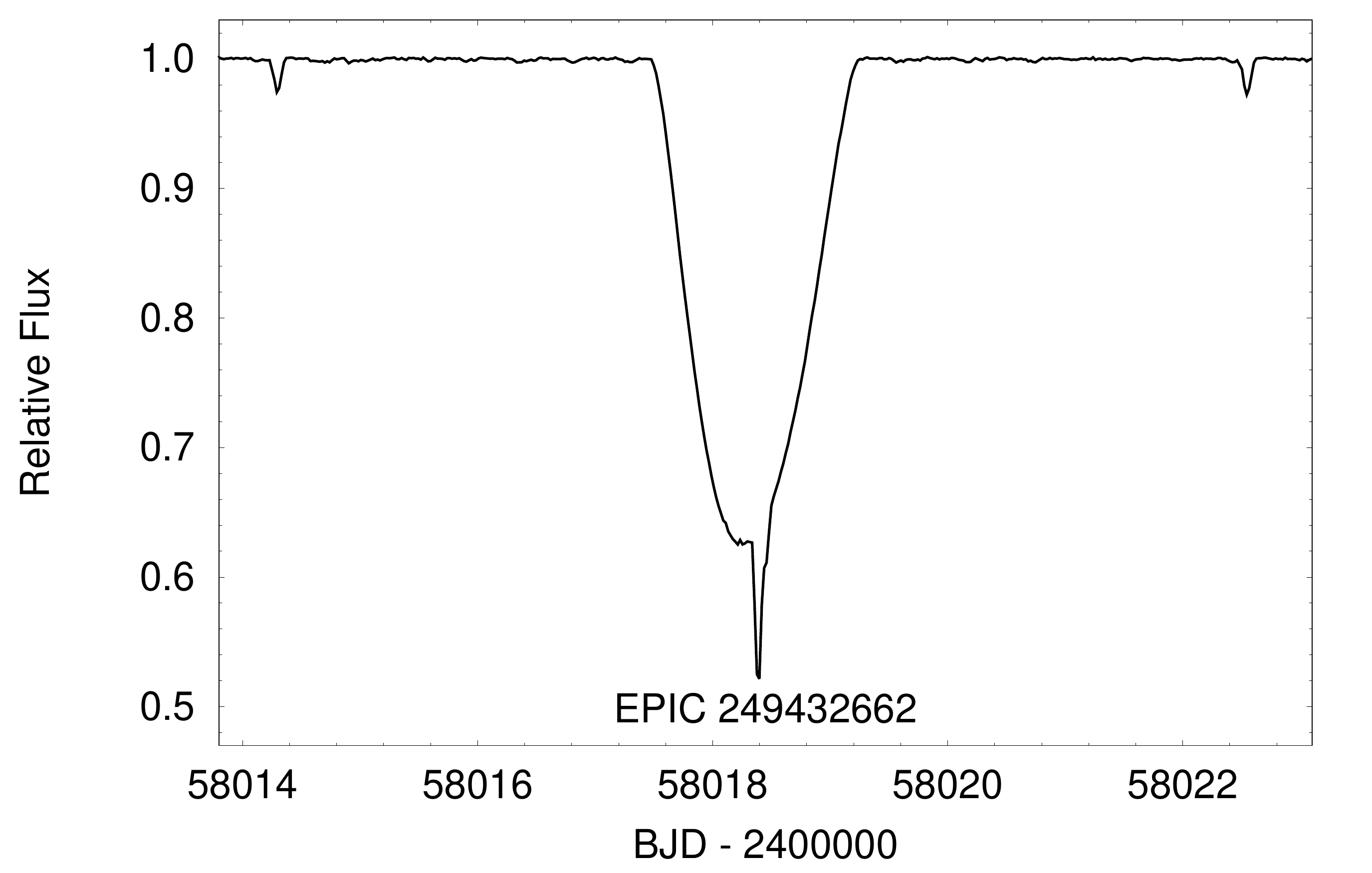}\includegraphics[width=4.5 cm]{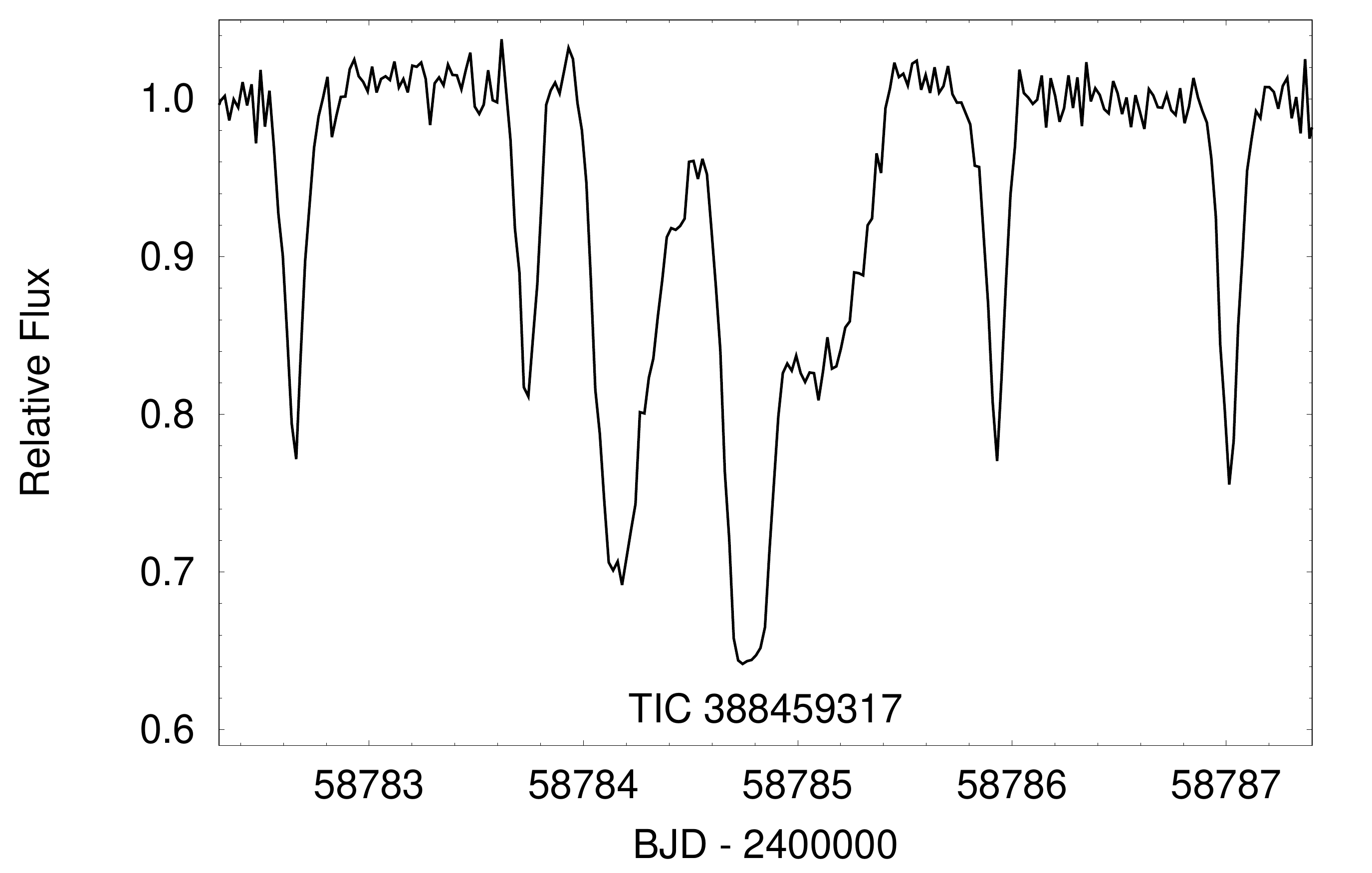}
\caption{Examples of third-body eclipses. Panels from left to right and from up to down are as follows: HD 181068 (KIC 5952403): a transit of the K dwarf-EB passing in front of the red giant third component. Due to the very similar effective temperatures of the giant and the K stars, regular eclipses are almost invisible during the central phase of the transit, with the exception of a sharp spike-like brightening around the middle of the third-body eclipse which is a regular secondary eclipse of the inner pair. On the other hand, tidal oscillations excited on the surface of the red giant by the K dwarfs with half of the synodic period are clearly visible. -- KIC~2856960: A typical set of extra eclipses of the `impossible triple star'; these have not yet been successfully modeled. -- KIC~5255552: Four separate extra eclipses from a single outer-orbit conjunction, which might be an indicator of the binary nature of the `third body'. -- KIC~6964043: Complex structured third-body eclipse. This EB no longer exhibits eclipses during the \textit{TESS} observations. -- EPIC~249432662: The discovery eclipse (and still only known eclipse) of the `Great Eclipser' in the \textit{K2} data. -- TIC~388459317: \textit{TESS}-observed complex-looking third-body eclipse.  Its morphology is somewhere between a `Great Eclipser'-like compound third-body eclipse and star-by-star separated third-body eclipses. \label{fig:3beclipses}}
\end{figure}

Third-body eclipses, in general, persist for longer durations, and have complex and variegated forms from event to event, even in the case of the same triple system. Moreover, their shape strongly depends on the photometric phase of the EB at the time of the middle of the extra eclipsing events. For example, if the middle of a third-body event (i.e., practically, the inferior or superior conjunction of the third star) coincides with an inner eclipse, one can observe a very long duration drop in flux caused by that star of the inner binary whose speed on the sky relative to the third star is small. In addition one can expect a second, sharp drop in flux in the middle of the long-duration event due the other binary star that moves into the opposite direction on the sky (see~Fig.~\ref{fig:3beclipses}). By contrast, when an extra event occurs near the quadrature of the inner EB, in many cases one will observe two separate extra eclipses that more or less resemble the usual, regular EB eclipses. Therefore, the presence of such kinds of extra eclipses in an EB light curve make it certain that the given target must be a hierarchical multiple (at least triple) system.  

The light curve of a triply eclipsing system carries several extra pieces of information relative to a usual EB light curve.  For example, as is well-known, the latter is scaled with the semimajor axis of the EB ($a_\mathrm{in}$).  By contrast, in the case of the third-body eclipses, a triple eclipser's light curve is scaled with both the outer semimajor axis ($a_\mathrm{out}$) and $a_\mathrm{in}$. As a consequence, one can obtain the ratio of the semimajor axes and, hence, the outer mass ratio ($q_\mathrm{out}$) can also be computed as long as we also know the inner and outer periods ($P_\mathrm{in,out}$). Several other parameters that can be determined in a purely geometric way are the inner mass ratio ($q_\mathrm{in}$), the multitude of orbital elements of both the inner and outer orbits. Even their relative orientations (i.e. the mutual inclination, $i_\mathrm{mut}$, and the difference of the observational nodes of the two orbital planes, $\Delta\Omega=\Omega_\mathrm{out}-\Omega_\mathrm{in}$) can also be inferred from a high quality light curve (see, e.g., Carter et al.~\cite{carteretal11}; Borkovits et al.~\cite{borkovitsetal13}; Marsh et al.~\cite{marshetal14}; Masuda et al.~\cite{masudaetal15}, Nemravov\'a et al.~\cite{nemravovaetal16} for detailed discussions).

Moreover, since the probability of third-body eclipses, in an analogous way to regular EB eclipses, decreases steeply with the size of the outer orbit (i.e, $\approxprop 1/a_{\rm out}$ or as $\approxprop 1/P_{\rm out}^{2/3}$), it is not at all surprising that the outer periods in most of the known cases are short.  In particular, triply eclipsing systems are in general not only ECHTs, but are tight enough for significant third-body perturbations. This fact opens the possibility for an independent, dynamical determination of many of the same orbital and mass parameters that are listed above. It really can be done {\it independently} through a combined LTTE+DE analysis of the ETV curve as was discussed in Sect.~\ref{sec:ETVs}. However, it is even more effective to carry out a combined photodynamical analysis, in which the light curve (sometimes multi-band); the ETVs; and, when available, RVs are analysed simultaneously.  Such an analysis naturally should be carried out in the time domain (instead of the old-fashioned method of using phase-folded lightcurves and/or RV curves), and also requires numerical integration of the orbital motion of each object, including the usual three- or more-body forces, the mutual tidal and stellar rotation terms and, additionally, the relativistic contributions as well.

Before the launch of the \textit{Kepler} space telescope no third-body eclipses had ever been observed and noted as such.   The first triply eclipsing ECHT ever discovered was KOI-126 (KIC~5897826). In the discovery paper Carter et al.~\cite{carteretal11} present their detailed photo-dynamical analysis, based on the first $\sim 14$ months of \textit{Kepler} observations, and on follow-up spectroscopy as well. KOI-126 consists of a $P_\mathrm{in}=1.77$\,d period inner pair of two very low mass M stars that are orbiting around a much more massive and brighter F-type star with a period of $P_\mathrm{out}=33.92$\,d.  (This remains the second shortest known outer period in a CHT.)  Due to the small, but non-zero, mutual inclinations of the two orbital planes ($i_\mathrm{mut}\approx8^\circ$) the orbits exhibit rapid orbital plane precession ($P_\mathrm{node}\approx2.73$\,yr) and therefore, eclipses cannot be observed at every conjunction. Most recently Yenawine et al.~\cite{yenawineetal21} carried out a revised photodynamical analysis with the use of the full \textit{Kepler} dataset and also with supplemental ground-based photometric and RV measurements. They were able to measure the masses and radii of the three stars with accuracies of 0.14\%-0.37\% and also made attempts to measure the apsidal motion constants of the M-dwarfs.

Almost in parallel with KOI-126, the triply eclipsing nature of HD~181068 (KIC~5952403) was also announced by Derekas et al.~\cite{derekasetal11}. In this triple system two K-dwarfs ($P_\mathrm{in}=0.91$\,d) revolve with an outer period of $P_\mathrm{out}=45.5$\,d around a red giant star. The K dwarfs suppress the solar-like oscillations of the giant star, but excite tidal oscillations on its surface with exactly half the synodic period of the binary, i.e., as seen from a corotating outer orbital frame (see upper left panel of Fig.~\ref{fig:3beclipses}). In this triple the two orbits are coplanar and circular (Borkovits et al.~\cite{borkovitsetal13}).

In addition to the two above mentioned systems, and not counting the case of the transiting circumbinary planets (see below), \textit{Kepler} observed clear third-star eclipses in nine additional systems during its prime mission (see Table~\ref{Tab:tripleeclipsers}).  Amongst these, KIC~2856960 displays the strangest and most mysterious extra eclipse patterns every 204 days (Armstrong et al.~\cite{armstrongetal12} -- see upper middle panel in Fig.~\ref{fig:3beclipses}). As Marsh et al.~\cite{marshetal14} pointed out, it is an `impossible triple star', as the complex structure and duration of the extra eclipses cannot be explained by eclipses from only a `simple' third star. They made efforts to model these eclipses in the context of both 2+1+1 and 2+2 quadruple star models, but they were unable to find physically realistic solutions. KIC~5255552, the second longest outer period triply eclipsing system (upper right panel of Fig.~\ref{fig:3beclipses}), might also be a quadruple star (see, e.g., Getley et al.~\cite{getleyetal20}).  Attempts to find reliable quadruple solutions for that system are in progress.  Another curious system is KIC~4150611 whose light curve, in addition to  third-body eclipses, also exhibits three sets of regular two-body eclipses, and stellar pulsations as well (Shibahashi \& Kurtz~\cite{shibahashikurtz12}). According to the detailed study of He\l miniak et al.~\cite{helminiaketal17} this target is most probably at least a quintuple stellar system. Note also KIC~2835289, a CHT, which exhibits third-body eclipses, even though the inner binary itself is only an ellipsoidal variable without eclipses (Conroy et al.~\cite{conroyetal14,conroyetal15}). Finally, on this topic, we note that Orosz~\cite{orosz15} reported one low-amplitude extra eclipse in the light curve of KIC~7670485. Hence, Borkovits et al.~\cite{borkovitsetal16} investigated the eclipse times of all the regular eclipses during the 4-year-long \textit{Kepler} data set, but did not find any ETVs. Therefore, the origin of that sole extra eclipse remains uncertain, and we do not consider this system to be a secure triply eclipsing CHT.

The much shorter observations of the \textit{K2} mission ($\sim$80 days) were less favourable in regard to the detection of new triply eclipsing triples. Thus, only two new ECHTs were discovered during this secondary phase of the \textit{Kepler} mission. HD~144548 was found in the Upper Scorpius OB association region. Due to its very short outer period ($P_\mathrm{out}=33.95$\,d) one campaign of \textit{K2} observations (C2), supported by a few ground-based RV measurements, was sufficient for a precise analysis of this triple (Alonso et al.~\cite{alonsoetal15}). In contrast to this, in the case of EPIC~249432662, because the outer period ($P_\mathrm{out}=188.4$\,d) is about twice the length of the \textit{K2} dataset, its \textit{Kepler} observations contained only a single, enormously large-amplitude, third-body eclipse (lower middle panel of Fig.~\ref{fig:3beclipses}).  In this system, dubbed the `Great Eclipser', the role of ground-based follow-up photometric and RV observations was critical, not only for a detailed analysis of the system, but simply for the determination of the outer period (Borkovits et al.~\cite{borkovitsetal19c}). In such a manner, the investigation of this ECHT, from the first detection of a solo third-body eclipse (and a few regular EB eclipses), through the organization of an international follow-up campaign, all the way to the end of the complex, photodynamical analysis, served as a prototype for the process now followed routinely in the case of the even shorter \textit{TESS} datasets.

As of 2021 November the photodynamical analyses of five \textit{TESS}-discovered triply eclipsing tiples have been published (see Table~\ref{Tab:tripleeclipsers}). Apart from TIC~278825952 which is located in the SCVZ of \textit{TESS} and, hence, it was almost continuously observed during Cycles 1 and 3 (Mitnyan et al.~\cite{mitnyanetal20}), the other four were seen by \textit{TESS} only during a few 25-27\,day-long sectors.  Therefore, inclusion of data sets of former, archival or, ongoing surveys such as SuperWASP (Pollaco et al.~\cite{pollaccoetal06}), ASAS-SN (Kochanek et al.~\cite{kochaneketal17}), ATLAS (Heinze et al.~\cite{heinzeetal18}) and KELT (Pepper et al.~\cite{pepperetal07}), as well as additional ground-based follow-up observations were crucial to their investigations (Borkovits et al.~\cite{borkovitsetal20a,borkovitsetal22}). Note also, we are aware of the discovery of some dozens of additional triply eclipsing triples in the \textit{TESS} data. Follow up observations (both photometric and spectroscopic) of a few of these are in progress. Therefore, it is certain that the number of the well-characterized triply eclipsing ECHTs will continue to grow in the near future. 

Despite the much smaller areas on the sky surveyed by the CoRoT spacecraft, with durations ranging from one to five months, Hajdu et al.~\cite{hajduetal17} reported the discovery of two additional triply eclipsing triples, CoRoTs~104079133 and 221664856. In the former system, from an analysis of the DE-dominated ETVs they were able to obtain a probable outer period of $P_\mathrm{out}\approx90$\,d. For the other system, however, no non-linear ETV was observable in the very short, $\sim1$\,month-long dataset. Therefore, similar to KIC~7670485 we do not count this system amongst the secure triply eclipsing triples and, therefore, do not list it in Table~\ref{Tab:tripleeclipsers}.

Most recently, Hajdu et al.~\cite{hajduetal22} identified two triply eclipsing ECHTs in the galactic bulge sample of the OGLE-IV survey. Both systems (OGLE-BLG-ECL-126114 and -187370) were formerly categorized as double EBs (Soszy\'nski et al.~\cite{soszynskietal16}).

There is one additional CHT for which the third-body eclipses were discovered entirely by ground-based studies. It is the naked-eye ellipsoidal variable b~Persei, which exhibits third-body eclipses every $\approx701$\,days. The third component was discovered by Hill et al.~\cite{hilletal976} through variations in the systemic RV of this single-lined (SB1) system. The outer subsystem was resolved with the Navy Precision Optical Interferometer (NPOI, Armstrong et al.~\cite{armstrongetal998}), and from these measurements Hummel et al.~\cite{hummeletal13} concluded that there is great chance of third-body eclipses.  Assuming that the extra fading event in the system observed at Baja Observatory (Hungary) on the night of 31 October/1 November 1995 (Heged\"us et al.~\cite{hegedusetal996}) was part of a third-body eclipse, they predicted the likely date of the forthcoming event which was then actually caught by the observers of The American Association of Variable Star Observers (AAVSO) in 2013 (precisely 9 outer orbital revolutions after the 1995 observations; Collins et al.~\cite{collinsetal14}). Subsequently, another third-body eclipse was also observed in 2015 (Collins~\cite{collins16}).

Considering the approximate number of triply eclipsing triples, instead of any attempt at a thorough realistic estimation, here we present only a very simple note that may serve as a guide.  As mentioned above, \textit{Kepler} observed secure third-body eclipses in 11 EBs during its $\sim4$\,yr-long prime mission. Large portions of the \textit{Kepler}-field were reobserved by \textit{TESS} in Sectors 14, 15, 40, 41 (and a smaller region was also measured in Sector 26). Thus, as of November 2021 2-4 sectors of \textit{TESS} lightcurves are available for the majority of the targets in the original \textit{Kepler}-field. In these \textit{TESS} sectors, five of the eleven triply eclipsing \textit{Kepler} ECHTs display {\it new} third-body eclipses (see the systems noted with asterisks in Table~\ref{Tab:tripleeclipsers}), including such longer outer-period ECHTs such as KIC~2835289 ($P_\mathrm{out}=755$\,d), and KIC~6543674 ($P_\mathrm{out}=1101$\,d).  Moreover, as far as we know, no new third-body eclipsers have been found by \textit{TESS} in the original \textit{Kepler}-field, irrespective of whether an object had been targeted by \textit{Kepler} or not. Therefore, this tends to suggest crudely that the scanning strategy of \textit{TESS} might have an efficiency of $\sim40$\% in finding triple eclipsing systems.

The first transiting circumbinary planet (TCBP) around an EB (Kepler-16) was discovered in 2011 (Doyle et al.~\cite{doyleetal11}). It was followed by the discovery of 12 additional TCBPs, including the multiplanet system of Kepler-47 (Orosz et al.~\cite{oroszetal12}). Most recently the discovery of the first two TCBPs in the \textit{TESS} sample was also announced (Kostov et al.~\cite{kostovetal20,kostovetal21a}). 

\begin{specialtable}
\caption{List of the known close binaries exhibiting third-body eclipses (in increasing order of the outer period)}
\label{Tab:tripleeclipsers}
\begin{tabular}{llll}
\toprule
Identifier & $P_1$ & $P_2$ & References \\
\midrule
KOI-126$^*$    & 1.77 & 33.92 & \cite{carteretal11} \\
HD~144548      & 1.63 & 33.95 & \cite{alonsoetal15} \\
HD~181068$^*$  & 0.91 & 45.47 & \cite{derekasetal11} \\
TIC~193993801  & 1.43 & 49.28 & \cite{borkovitsetal22} \\
TIC~388459317  & 2.18 & 88.86 & \cite{borkovitsetal22} \\
CoRoT~104079133& 2.76 & 90(?) & \cite{hajduetal17} \\
KIC~4150611$^*$& 1.52 & 94.2  & \cite{shibahashikurtz12}, \cite{helminiaketal17}\\
OGLE-BLG-ECL-126114&6.65&105.3& \cite{hajduetal22} \\
TIC~209409435  & 5.72 & 121.9 & \cite{borkovitsetal20a} \\
TIC~52041148   & 1.79 & 177.1 & \cite{borkovitsetal22} \\
EPIC~249432662 & 8.19 & 188.4 & \cite{borkovitsetal19c} \\
KIC~2856960    & 0.26 & 204.8 & \cite{armstrongetal12}, \cite{marshetal14}\\
KIC~7668648    & 27.83& 204.8 & \cite{borkovitsetal15}, \cite{orosz15} \\
TIC~278825952  & 4.78 & 235.5 & \cite{mitnyanetal20} \\ 
KIC~6964043    & 10.73& 239.1 & \cite{borkovitsetal15}\\
KIC~7289157    & 5.27 & 243.4 & \cite{borkovitsetal15}, \cite{orosz15} \\
OGLE-BLG-ECL-187370&11.96&280.5& \cite{hajduetal22} \\
KIC~9007918    & 1.39 & 470.9 & \cite{borkovitsetal16}\\
b~Persei       & 1.52 & 704.5 & \cite{collinsetal14} \\
KIC~2835289$^*$& 0.86 & 755   & \cite{conroyetal14} \\
KIC~5255552    & 32.47& 862.1 & \cite{borkovitsetal15} \\ 
KIC~6543674$^*$& 2.39 & 1101.4& \cite{borkovitsetal15}, \cite{masudaetal15} \\
\bottomrule
\end{tabular}

\small
$^*$: These stars were identified in the original \textit{Kepler}-field as triply eclipsing triples, and new third-body eclipses were also observed with \textit{TESS}.
\end{specialtable}

\subsubsection{Double (multiple) EBs}

In contrast to third-body eclipses, regular eclipses of two (and in a few exceptional cases, three) EBs blended in the same light curve do not show any obvious connection to each other, and therefore, in most cases, can be readily disentangled (see, e.g., Powell et al.~\cite{powelletal21}).  As a direct consequence, in this case, the light curve itself is not sufficient to decide whether the blended EBs have any physical connection to each other (i.e. if they form a 2+2 quadruple system), or not. This is especially true for \textit{TESS} due to its large pixel size. In that case, however, if both EBs of a blended light curve display ETVs with the same period, then one can be certain that the two EBs form a bound, compact 2+2 type quadruple system. In contrast to the triply eclipsing triples, a few double EBs were already known before the era of \textit{Kepler}. 

The doubly eclipsing nature of the relatively bright, eccentric EB, V994 Her, was first reported by Lee et al.~\cite{leeetal08}. Then Zasche \& Uhla\v r~\cite{zascheuhlar13} pointed out that the ETV curves of the two EBs ($P_\mathrm{in,A}=2.08$\,d; $P_\mathrm{in,B}=1.42$\,d) show anticorrelated behaviour, robustly implying that they form a relatively close 2+2 quadruple system. Finally, with the use of new additional eclipse times Zasche \& Uhla\v r~\cite{zascheuhlar16} determined the outer period to be $P_\mathrm{out}=1063$\,d.

Despite the fact that they may have been discovered relatively easily, compact double EBs are actually fairly rare objects.  Recently discovered systems include only EPIC~220204960 ($P_\mathrm{in,A}=13.27$\,d; $P_\mathrm{in, B}=14.42$\,d; $P_\mathrm{out}=300-500$\,d -- Rappaport et al.~\cite{rappaportetal17}), TIC~454140642 ($P_\mathrm{in,A}=13.62$\,d; $P_\mathrm{in, B}=10.39$\,d; $P_\mathrm{out}=432$\,d -- Kostov et al.~\cite{kostovetal21b}), BG~Ind ($P_\mathrm{in,A}=1.46$\,d; $P_\mathrm{in, B}=0.53$\,d; $P_\mathrm{out}=721$\,d -- Borkovits et al.~\cite{borkovitsetal21}) and TIC~278956474 ($P_\mathrm{in,A}=5.49$\,d; $P_\mathrm{in, B}=5.67$\,d; $P_\mathrm{out}=858$\,d -- Rowden et al.~\cite{rowdenetal20}). Note also, the intriguing case of BU~CMi. For this double EB Volkov et al.~\cite{volkovetal21} recently reported an outer period of $P_\mathrm{out}=6.54$\,yr, but at the same time, Jayaraman et al.~\cite{jayaramanetal21} have found a much shorter, record-holding outer period of $P_\mathrm{out}\approx120$\,d.
 
Besides the four or five compact quadruple systems listed above, there is only one further 2+2 system for which the outer period is less than 1000\,days. It is the current record-holder 2+2 system VW~LMi with outer period of $P_\mathrm{out}=355$\,days (see Sect.~\ref{sec:RVs}).  It consists of a W~UMa-type overcontact EB ($P_\mathrm{in,A}=0.48$\,d) and a non-eclipsing second pair with $P_\mathrm{in,B}=7.93$\,d. There are also two more compact systems, however, which currently are known as ECHTs, but some authors claim that the third components are actual close binaries. These are IU~Aur ($P_\mathrm{in}=1.83$\,d; $P_\mathrm{out}=293.3$\,d -- Drechsel et al.~\cite{drechseletal994}; \"Ozdemir et al.~\cite{ozdemiretal03}) and V1200~Cen ($P_\mathrm{in}=2.48$\,d; $P_\mathrm{out}=180.4$\,d -- Marcadon et al.~\cite{marcadonetal20}).

Finally we note the interesting case of OGLE-LMC-ECL-15674 which is a double EB ($P_\mathrm{in,A}=1.43$\,d; $P_\mathrm{in,B}=1.39$\,d) in the Large Magellanic Cloud. Analysing observations from 1992 to 2016 Hong et al.~\cite{hongetal18} found evidences for eclipse depth variations in the EB's light curves, from which they inferred $\approx6-8^\circ$ inclination variations in both EBs. If this is the consequence of a non-aligned outer orbit relative to the orbits of the two inner EBs then one may expect an outer period around a hundred days. Therefore, this system is worthy of future detailed studies.

\subsection{Eclipse depth variations -- disappearing eclipses}
\label{sec:depthvar} 

The first evidence of eclipse depth variations (EDVs) in EBs was reported by Hall \cite{hall969}, for RW Persei, and by Mayer \cite{mayer971}, for IU Aurigae (we list all EBs with reported EDVs in our Galaxy, in Table~\ref{Tab:EBprec}).  Interestingly, in the case of the latter system, Mayer immediately interpreted this fact as due to inclination variation caused by the ($P_\mathrm{out}=293.3$\,day-period) tertiary, discovered a few years earlier through ETV analysis. While for RW Per, nodal motion caused by a third component as a possible explanation of the EDVs was first proposed only 22 years later by Shaefer \& Fried \cite{schaeferfried991} (this hypothesis was later revised by Olson et al. \cite{olsonetal992}).

The total disappearance of the eclipses in a formerly eclipsing system (AY Mus) was first discovered by S\"oderhjelm \cite{soderhjelm974,soderhjelm975} who, from the rate of the EDVs, estimated the third-body period to be about 0.5 -- 3\,yrs.  Another former EB that no longer exhibits eclipses is SS Lac. The disappearance of its eclipses first was reported by Zakirov \& Azimov~\cite{zakirovazimov990}. The system was then intensively studied over the ensuing decade. The reason for this great interest was that, even though Lehmann~\cite{lehmann991} and Mossakovskaya~\cite{mossakovskaya993}, pointed out the continuous EDVs during the first half of the 20th century and, therefore, concluded that the most probable explanation was orbital plane precession due to a third body, Schiller \& Milone~\cite{schillermilone996} proposed an alternative hypothesis. According to their explanation the binary SS~Lac, which was supposed to be a member of the Open Cluster NGC~7209, might have been disrupted due to a close encounter with another cluster star member. This latter hypothesis was based on the non-detection of RV variations in the former EB. Later, however, the same authors carried out new and more careful spectroscopic and photometric analysis, and revised their former assumption, accepting the third-body hypothesis (Milone et al.~\cite{miloneetal00}). Finally, the spectroscopic analysis of Torres \& Stefanik~\cite{torresstefanik00} did reveal that the system is really intact and contains a third star with an outer period of $P_\mathrm{out}=679$\,d. Torres \cite{torres01} calculated the nodal period to be about $P_\mathrm{prec}\sim600$\,yr. 

Another remarkable system is V907~Sco which `is unique among all known EB stars because its eclipses have turned on and off twice within modern history' (Lacy et al. \cite{lacyetal999}). By the use of a 11\,yr-long spectroscopic data train of this double-lined spectroscopic binary (SB2) they found evidence for the presence of a $P_\mathrm{out}=99.3$\,d period third stellar component. They estimated a nodal period of $P_\mathrm{node}\sim68$\,yr, and predicted that the eclipses which stopped between 1980.6 and 1988.2 will return around $2030\pm5$. In this regard, note that the \textit{TESS} spacecraft observed V907~Sco in Sectors 13 (Year 1, Cycle 1) and 39 (Cycle 3). During the Sector 13 observations (19 June -- 18 July 2019) the EB exhibits $\sim$$0.005$\,mag amplitude ellipsoidal variations and, moreover, $~0.001$\,mag amplitude, very short grazing eclipses that are visible in the middle of every second ellipsoidal light minimum. Two years later, however, the Sector 39 (26 May -- 24 June 2021) light curve, besides similar ellipsoidal variations, exhibits clear $\sim$$0.05$\,mag primary and slightly shallower secondary eclipses. Therefore, one can conclude that the eclipses of V907~Scorpii have returned about ten years earlier than they were expected.

EDVs in the opposite direction were detected in HS~Hya. Zasche \& Paschke~\cite{zaschepaschke12} found that the inclination of the system decreased by more than $15^\circ$ between 1964 and 2008. They calculated a $\sim631$\,yr nodal period caused by the perturbations of a $\sim190$\,d period, distant third companion, and predicted the disappearance of the eclipses around $\sim$2022. Most recently Davenport et al. \cite{davenportetal21} investigated the \textit{TESS} observations of HS~Hya. They detected $0.00173\pm0.00007$\,mag amplitude primary eclipses in the March 2019 (Sector 9) data, while no eclipses were found in the February 2021 (Sector 35) observations. They predict the recurrence of the eclipses in 2195.

\begin{specialtable} 
 \caption{EB's with eclipse depth variations (`EDVs') in our Milky Way. The first section (above the KIC systems) tabulates those systems where the detection of the EDVs was based on ground-based photometric data only. The middle part represents the EBs whose EDVs were observed by the \textit{Kepler} spacecraft. In the last two systems, EDVs were detected due to \textit{TESS} photometry.}
 \label{Tab:EBprec}
 \begin{tabular}{@{}cccccc}
  \toprule
System & $P_\mathrm{in}$ & $P_\mathrm{out}$ & $P_\mathrm{node}$ & current & References \\
       & [day] & [day] & [yr] & status & \\
\midrule
IU Aur   & 1.8115  & 293.3 & 330 & d & \cite{ozdemiretal03} \\
QX Cas   & 6.0047  &  ?    &  ?  & n & \cite{guinanetal12} \\
V685 Cen & 1.1910  &  ?    &  ?  & d?& \cite{mayeretal04} \\
AH Cep   & 1.7747  &  ?    &  ?  & i & \cite{drechseletal989} \\
V699 Cyg & 3.1031  &  ?    &  ?  & n & \cite{azimovzakirov991} \\
SV Gem   & 4.0061  &  ?    &  ?  & n & \cite{guilbaultetal01} \\
HS Hya   & 1.5680  & 190   & 631 & n & \cite{zaschepaschke12} \\
SS Lac   & 14.4162 & 679   & 600 & n & \cite{torres01} \\
AY Mus   & 3.2055  &  ?    &  ?  & n & \cite{soderhjelm974} \\ 
RW Per   & 13.1989 &  ?    &  ?  & d & \cite{olsonetal992} \\
V907 Sco & 3.7763  &  99.3 & 68? & i & \cite{lacyetal999} \\
\midrule
KIC 4769799 & 21.9286 & 1231 & 826  & n & \cite{borkovitsetal15} \\
KIC 5003117 & 37.6094 & 2128 & 1484 & n & \cite{borkovitsetal15} \\
KIC 5255552 & 32.4486 & 883  & 140  & i & \cite{borkovitsetal15} \\
KIC 5653126 & 38.4923 & 968  & 157  & i & \cite{borkovitsetal15} \\
KIC 5731312 &  7.9464 & 911  & 1013 & d & \cite{borkovitsetal15} \\
KIC 5771589 & 10.7382 & 113  &  7.5 & i & \cite{borkovitsetal15} \\
KIC 5897826 &  1.7671 &33.92 &  2.7 & ? & \cite{carteretal11,yenawineetal21} \\
KIC 6964043 & 10.7255 & 239.1&  26  & n & \cite{borkovitsetal15} \\
KIC 7289157 &  5.2665 & 243.4&  80  & n & \cite{borkovitsetal15} \\
KIC 7668648 & 27.8256 & 204.8&  25  & i & \cite{borkovitsetal15} \\
KIC 7670617 & 24.7032 & 3304 & 1678 & n & \cite{borkovitsetal15} \\
KIC 7955301 & 15.3278 & 209.1& 19.2 & n & \cite{borkovitsetal15} \\
KIC 8023317 & 16.5791 & 607  &  633 & c & \cite{borkovitsetal15} \\
KIC 8143170 & 28.7868 & 1710 &  890 & c & \cite{borkovitsetal15} \\
KIC 8938628 & 6.8622  & 388.6&  170 & n & \cite{borkovitsetal15} \\
KIC 9715925 & 6.3083  & 736  & 1163 & i & \cite{borkovitsetal15} \\
KIC 9963009 & 40.0716 & 3770 & 2703 & c?& \cite{borkovitsetal15} \\
KIC 10268809 &24.7084 & 7000 & 3333 & n?& \cite{borkovitsetal15} \\
KIC 10319590 &21.3212 & 452  &  110 & n & \cite{borkovitsetal15,orosz15} \\
KIC 10666242 &81.2455 & ?    &    ? & ? & \cite{borkovitsetal16} \\
HIP 41431   &  2.9329 & 58.9 &   11 & d & \cite{borkovitsetal19b} \\
\midrule
TIC 167692429 & 10.2648 & 331.5 & 70 & d & \cite{borkovitsetal20b} \\
VV Ori   & 1.4854  &  ?    &  ?  & d & \cite{southworthetal21} \\
\bottomrule
\end{tabular}

\small
(In column `current status':) d -- decreasing; i -- increasing; c -- constant eclipse depths; n -- no eclipses visible

\end{specialtable}

\begin{figure}[H]
\includegraphics[width=4.5 cm]{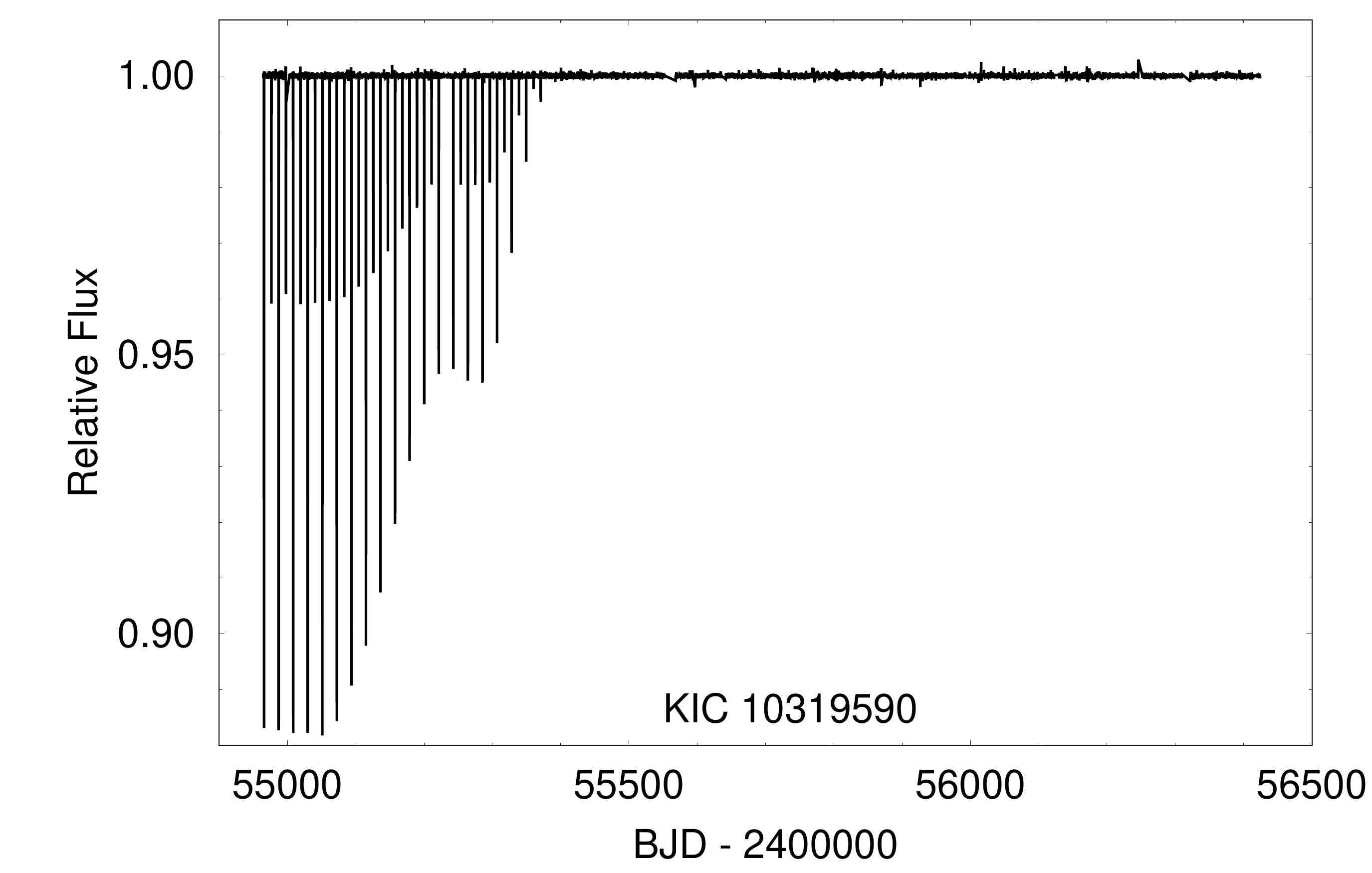}\includegraphics[width=4.5 cm]{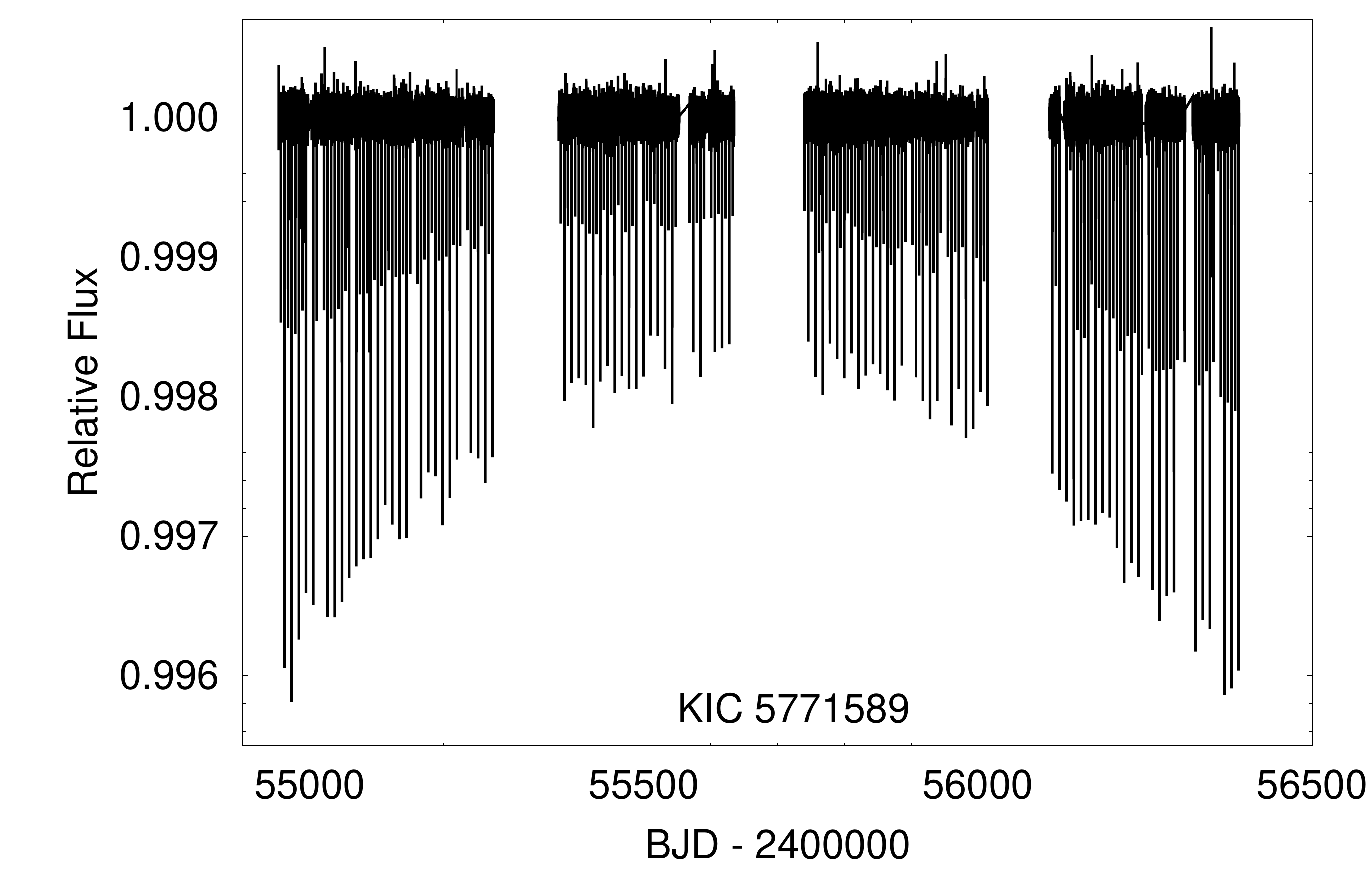}\includegraphics[width=4.5 cm]{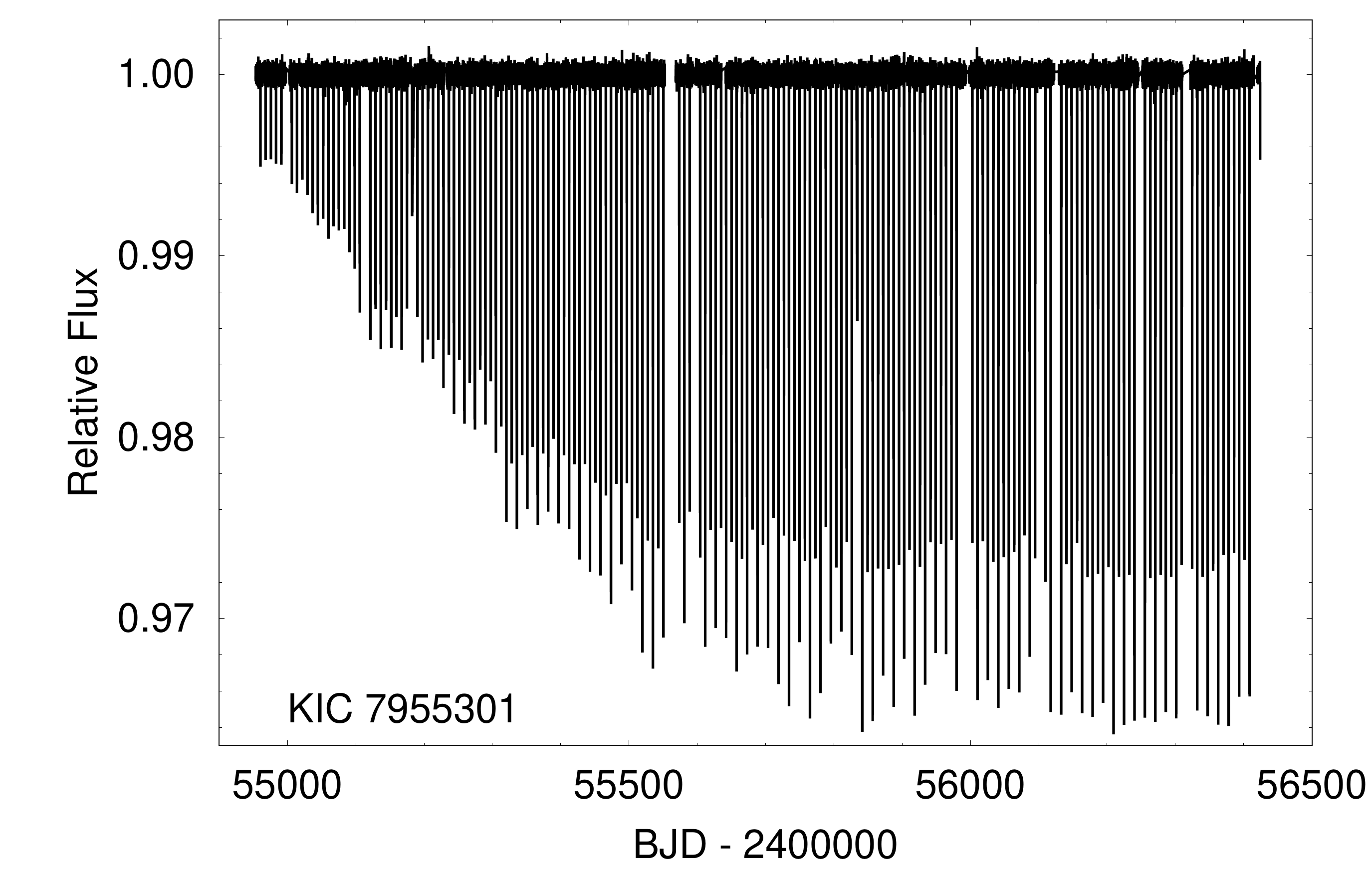}
\includegraphics[width=4.5 cm]{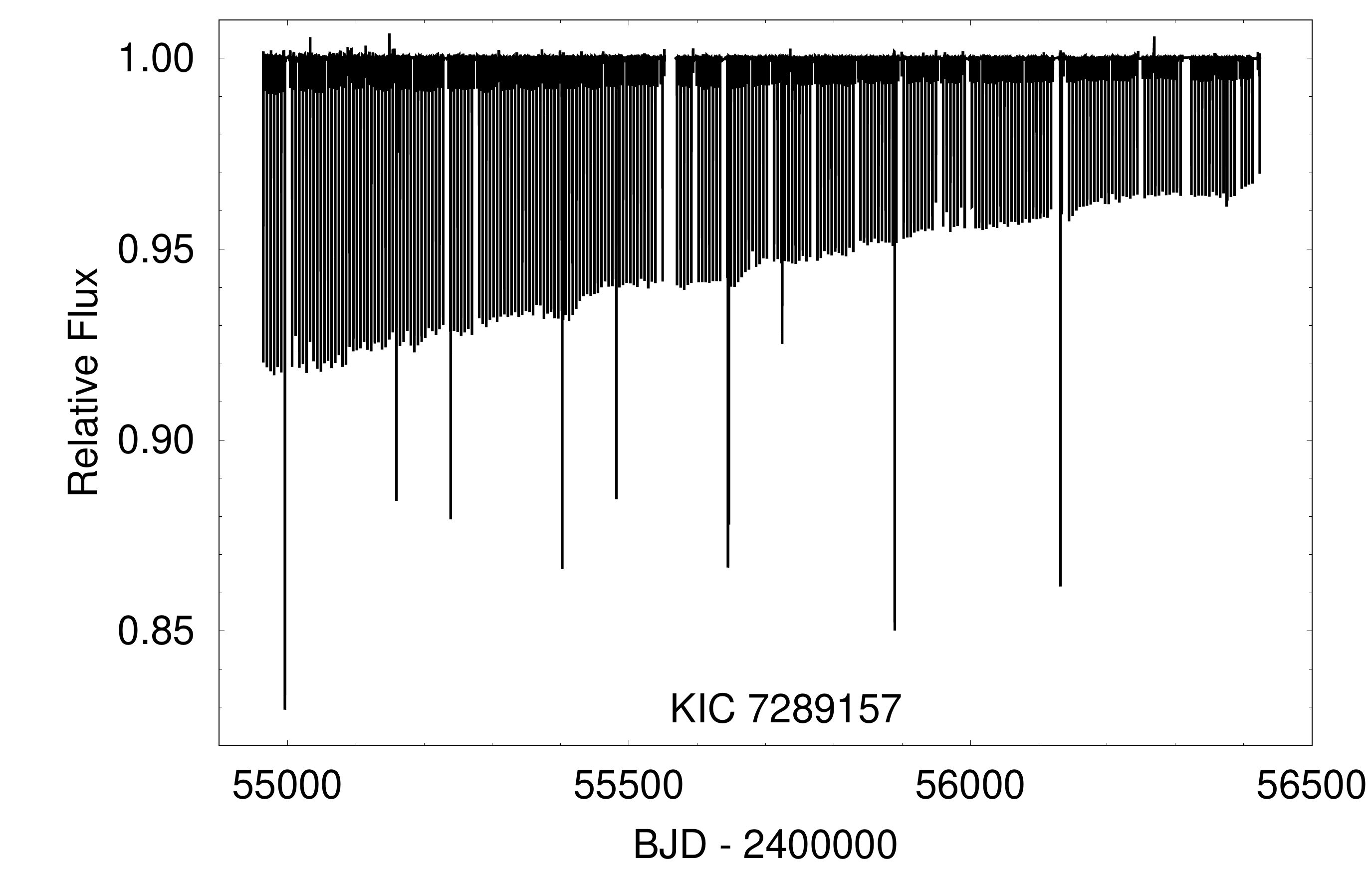}\includegraphics[width=4.5 cm]{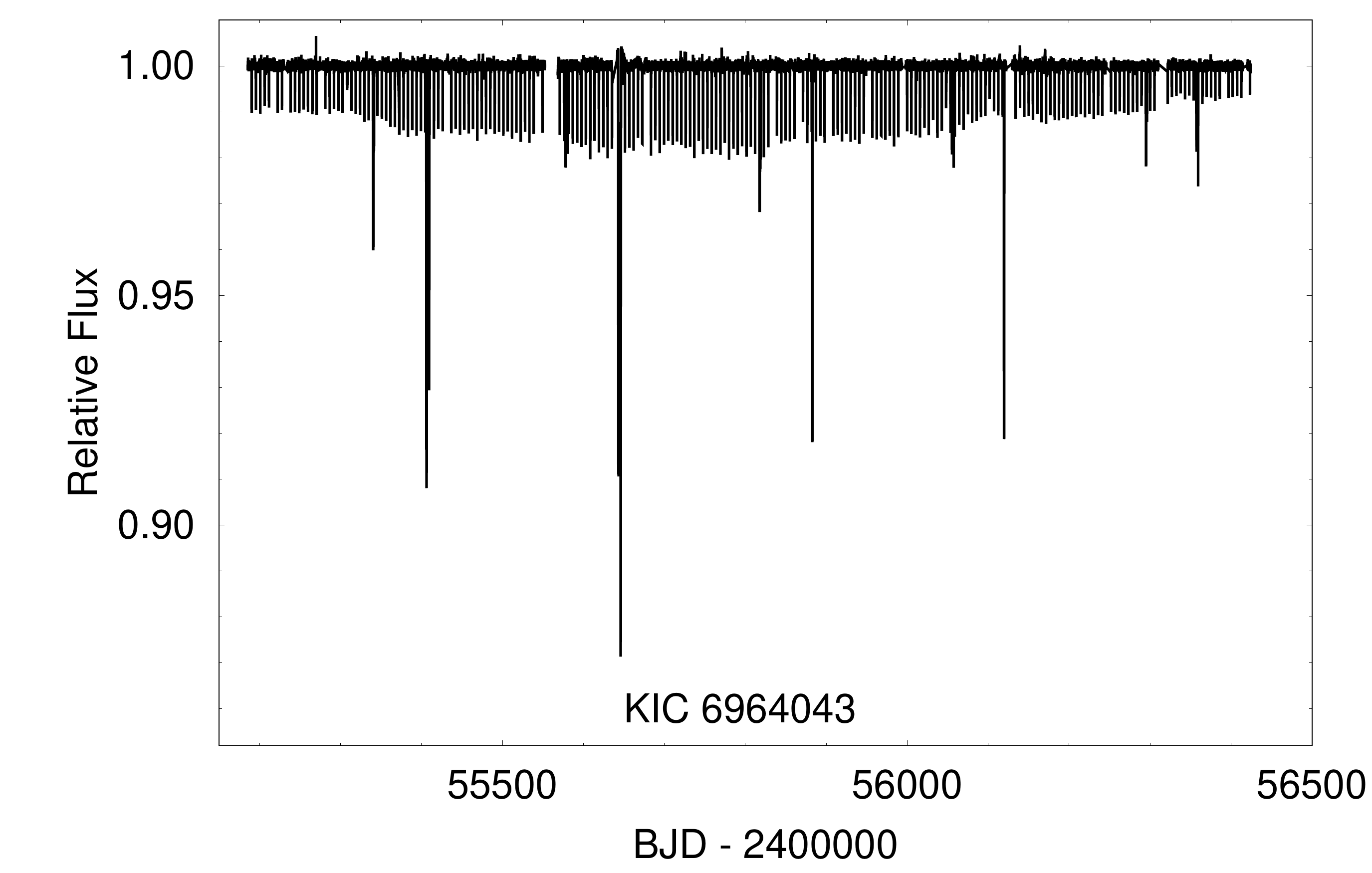}\includegraphics[width=4.5 cm]{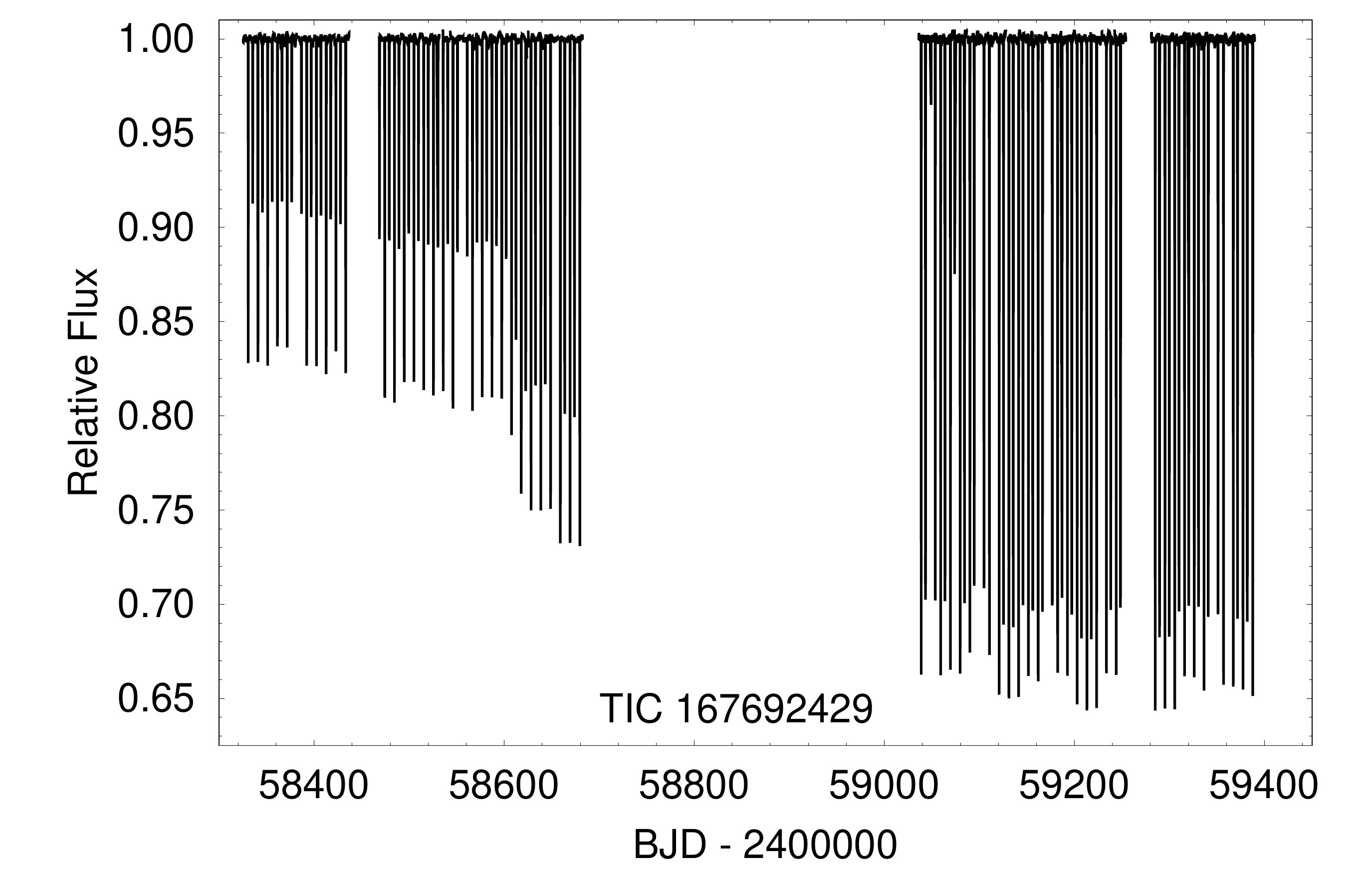}
\caption{Examples for EDVs amongst ECHTs in the \textit{Kepler} and \textit{TESS} fields. In the case of KIC 10319590 (upper left panel) and TIC 167692429 (lower right panel) the periastron passages of the outer orbits can also be identified as jumps that break the otherwise smooth trends. The sections of practically constant eclipse depths at the right ends of the light curves of KIC 7955301 (upper right) and TIC 167692429 (lower right) result from the fact that there the eclipses became central (i.e., total and annular ones). Finally, note the extra spike-like dips for KICs 7289157 (lower left) and 6964043 (lower middle) which are third-body eclipses.}
\label{fig:EDVs}
\end{figure}

Prior to the \textit{Kepler} era only 11 EBs were known to have EDVs.  The four-year-long \textit{Kepler} mission, with its quasi-continuous photometry, nearly tripled this number.  Perhaps the most remarkable one is the case of KIC~10319590, where the EB showed eclipses with constant depths during the first 200 days of the \textit{Kepler} observations (see upper left panel of Fig.~\ref{fig:EDVs}). Then, however, the depths began to decrease rapidly, and the eclipses disappeared completely after $\sim420$\,days (Rappaport et al.~\cite{rappaportetal13}).   Note that this EB, together with the \textit{Kepler} EBs listed in Table~\ref{Tab:EBprec}, also exhibited clear third-body forced dynamical effects in their ETVs (see lower left panel of Fig~\ref{fig:ETVs}); this was reported for the first time by Slawson et al.~\cite{slawsonetal11}. After the preliminary ETV analysis of Rappaport et al.~\cite{rappaportetal13}, the more sophisticated analysis of Borkovits et al.~\cite{borkovitsetal15} found that the $P_\mathrm{in}=21.3$\,d period inner binary has a $P_\mathrm{out}=452$\,d period third stellar companion which revolves on an inclined orbit ($i_\mathrm{mut}=40^\circ$). They estimated the nodal period to be $P_\mathrm{node}\approx100$\,yr. These results were essentially confirmed by the spectro-photodynamical analysis of Orosz~\cite{orosz15} which made use of spectra obtained with the echelle spectrograph on the Kitt Peak 4m telescope. Thus far, unfortunately, these latter results appear only in a short note cited in a conference proceedings.

Despite the much larger sky coverage of the \textit{TESS} spacecraft, its very short observing windows for most targets are unfavourable for the detection of EDVs.  On the other hand, since it reobserves the vast majority of the previously known EBs it makes it possible to check the status of many of the formerly known EBs with EDVs. We noted above the findings of the new \textit{TESS} observations of HS~Hya and V907~Sco. In the fifth column of Table~\ref{Tab:EBprec} we give the current state of these EBs after revisiting their short \textit{TESS} data series (at least if they are available). As one can see, a substantial portion of EBs with EDVs detected by \textit{Kepler} did not exhibit eclipses during the 2019--2021 year reobservations of the prime \textit{Kepler}-field. For example, one of these no-longer-eclipsing \textit{Kepler} EBs is KIC~7955301 (upper right panel of Fig.~\ref{fig:EDVs}), which consists of a red giant tertiary (Gaulme et al.~\cite{gaulmeetal13}). According to a recent spectro-photodynamical analysis to be published soon, the small mutual inclination of $i_\mathrm{mut}\approx6^\circ$ between the inner and outer orbits forces $P_\mathrm{node}=19.2$\,yr-period precession cycles with $\sim$7.3\,yr-long sessions of regular eclipses. The next eclipsing session is expected to begin in 2028 (Gaulme et al.~\cite{gaulmeetal22}).

To this point the \textit{TESS} observations have led to the discovery of previously unknown EDVs in two EBs. One of them is the well-known, naked-eye EB VV~Orionis (which is also a $\beta$~Cephei-type pulsator). Formerly this system exhibited total (U-shaped) eclipses (e.g., Chambliss \& Leung~\cite{chamblissleung982}); however, as was pointed out by Southworth et al.~\cite{southworthetal21} the eclipses in the \textit{TESS} light curves are clearly V-shaped, i.e., partial ones.  Reinvestigating all the available historical observations, these authors concluded that the system was seen  exactly edge-on in the 1950s or 1960s, while nowadays its inclination is decreasing at a rate of $0.2^\circ\,\mathrm{yr}^{-1}$. This is most probably due to an unknown, third stellar component. We also note that, well before the recent findings of EDVs, a $P_\mathrm{out}=120$\,d period third star was tentatively identified from RV data (see, e.g., Daniel~\cite{daniel915}; Chambliss~\cite{chambliss983}); however, a careful detailed analysis by Van Hamme \& Wilson~\cite{vanhammewilson07} has found no evidence of its existence (see also Terrell et al.~\cite{terrelletal07}). 

The other EB with EDV identified in \textit{TESS} data is TIC 167692429 (lower right panel of Fig.~\ref{fig:EDVs}). In this system the inner binary is moderately inclined ($i_\mathrm{mut}=27^\circ$) with respect to the outer orbit. This fact manifests itself in rapid eclipse depth variations, as a consequence of the orbital plane precession ($P_\mathrm{node}\approx70$\,yr). During such a precession cycle, the inner binary exhibits eclipses during two $\sim$$11$\,yr-long intervals (Borkovits et al.~\cite{borkovitsetal20b}).

Regarding extragalactic EBs, EDVs in some EBs in the Magellanic Clouds observed as part of the OGLE-III survey were reported by Graczyk et al.~\cite{graczyketal11}. The first targeted analysis was carried out by Zasche \& Wolf~\cite{zaschewolf13}, who investigated the inclination variations of the $P_\mathrm{in}=1.26$\,d period detached EB MACHO 82.8043.171. They analysed 9 sets of MACHO, OGLE, and their own light curves obtained at different epochs, and determined (amongst other parameters) the inclination variation rate and made estimations for the parameters of the third star. Recently Jury\v{s}ek et al.~\cite{juryseketal18} investigated the OGLE-III data to search for EBs that exhibit EDVs. In such a manner, they found 58 new CHT candidates, from which they were able to carry out detailed analysis for eight systems.
 
Finally, we note that apsidal motion in an EB can also produce EDVs, in addition to being caused by orbital plane precession. This is so because, in an eccentric system, apart from an exactly edge-on configuration (i.e., $i=90^\circ$), the impact parameter on the sky between the two stars at conjunctions depends on the argument of periastron ($\omega$) as well. This effect, however, has much lesser observational relevance and, therefore, in most cases is overlooked.

\subsection{Astrometry/Interferometry}
\label{sec:astrometry}

Astrometry is the most classic `ancient' branch of astronomy and, moreover, astrometry of binary stars is one of the oldest branches within observational astrometry.  Previously, though, astrometric measurements had a fairly minor role in the study of the most compact triple and multiple stellar systems.  This is quite natural due to the very small angular separations of most of these systems, which correspond to similarly small angular wobbles caused by the sky-projected motion of any of the visually observable components of a CHT along its orbit around the CM of the system.  As an early example, we refer to the thorough work of Bachmann \& Hershey~\cite{bachmannhershey975} who carried out an orbital analysis of Algol AB, C (i.e., the outer orbit of this emblematic CHT) with the simultaneous use of astrometric, photometric (ETV), and spectroscopic (RV) data. From the astrometric side they utilized 40 measured positions between epochs 1924.00 and 1973.00. They determined the orbital elements and period of the outer orbit with the use of both separate datasets and their combinations. 

The advent of different interferometric techniques reaching higher and higher angular resolutions, however, resulted in some important advances in this field as well. As a first step, Speckle Interferometry made it possible to resolve the outer components of some CHTs starting as early as the 1970s.  For the first time, none other than Algol C, the third star in the $\beta$ Persei system was resolved in such a manner in 1973 (Labeyrie et al.~\cite{labeyrieetal974}). With the use of the eight successful speckle position measurements between 1973.4500 and 1977.0867 Bonneau~\cite{bonneau979} were able to determine the observable inclination ($i_\mathrm{out}$) and node ($\Omega_\mathrm{out}$) with a few percent accuracy (apart from the usual ambiguity of $180^\circ$ in the node) and also obtained the stellar masses of the two components of the outer binary ($m_\mathrm{AB}$ -- i.e., the total mass of the inner EB, and $m_\mathrm{C}$) with $\approx10$\% accuracy. These results were later refined by Pan et al.~\cite{panetal993}.

Further on the topic of Algol, Lestrade et al.~\cite{lestradeetal993}, in their epochal work, report the first resolved motions of the close EB in Algol in the radio domain with the use of Very Long Baseline Interferometry (VLBI).  They were able to detect the signal and movement of the radio emitting evolved, secondary component, Algol B. In this way, they measured four precise astrometric positions of component B (two near the first quadrature of the EB, and the others near the second quadrature). They thereby determined the node of the inner orbit ($\Omega_\mathrm{in}$).\footnote{Previously, Rudy~\cite{rudy979} and Kemp et al.~\cite{kempetal981,kempetal983} had obtained very similar values with optical polarimetric observations, though these results were much less certain.}

Resolving the close inner binary of a CHT leads to the possibility of determining its astrometric orbit, especially the longitude of the node ($\Omega_\mathrm{in}$).  This, in turn, opened the door to determining such an important dynamical and evolutionary parameter of a triple system as the mutual inclination ($i_\mathrm{mut}$) of the inner and outer orbital planes. This determination depends on the simple spherical geometric relation (see., e.g., Borkovits et al.~\cite{borkovitsetal03})
\begin{equation}
\cos i_\mathrm{mut}=\cos i_\mathrm{in}\cos i_\mathrm{out}+\sin i_\mathrm{in}\sin i_\mathrm{out}\cos(\Omega_\mathrm{out}-\Omega_\mathrm{in}).
\label{eq:imut}
\end{equation}
In the case of Algol the VLBI results led to a nearly perpendicular configuration, contradicting to the results of S\"oderhjelm~\cite{soderhjelm980}, who, from the numerically very similar values of the inner and outer inclinations, and the lack of any EDVs in the last century, arrived at the conclusion that this triple must be coplanar. 

With the advance of the observational techniques, by the end of the last millenium long-baseline optical interferometers utilizing even modest size telescopes became able to achieve milliarcsecond (mas) precision, or, in exception cases, even submilliarcsecond-level global astrometric precision. Thus, systematic determinations of the mutual inclinations in relatively bright and nearby hierarchical triple stellar systems have begun. For example, as a first attempt, with the use of NPOI, Hummel et al.~\cite{hummeletal03} resolved all the three components of $\eta$~Vir ($P_\mathrm{in}=71.79$\,d; $P_\mathrm{out}=4774$\,d), and found a mutual inclination of $i_\mathrm{mut}=30.8^\circ$. Shortly thereafter, with the use of the Palomar Testbed Interferometer (PTI, Colavita et al.~\cite{colavitaetal999}) systematically determined the mutual inclinations of several hierarchical triple systems with a high-resolution differential astrometry technique described in Lane \& Muterspaugh~\cite{lanemuterspaugh04}. Here we mention two of them, the eclipsing system V819~Her ($P_\mathrm{in}=2.23$\,d; $P_\mathrm{out}=2020$\,d; $i_\mathrm{mut}=23.6\pm4.9^\circ$ -- Muterspaugh et al.~\cite{muterspaughetal06a}, and $\kappa$~Peg ($P_\mathrm{in}=5.97$\,d; $P_\mathrm{out}=4237$\,d; $i_\mathrm{mut}=43.8\pm3.0^\circ$ -- Muterspaugh et al.~\cite{muterspaughetal06b}). Note however that, strictly speaking, none of these triples can be categorized as CHTs because their outer periods are longer than 1000\,days. On the other hand, the inner periods in the last two highlighted systems are typical for (E)CHTs, demonstrating that there are no technical barriers to the interferometric determination of mutual inclinations in appropriately bright CHTs.

Returning back to Algol, taking the inclinations and node of the inner orbit from S\"oderhjelm~\cite{soderhjelm980} and Lestrade et al.~\cite{lestradeetal993}, respectively, and the same orbital elements of the outer orbit from Pan et al.~\cite{panetal993}, Kiseleva et al.~\cite{kiselevaetal998} calculated a mutual inclination of $i_\mathrm{mut}\approx100^\circ$. If this value were correct, however, the inclination of the EB would have changed by $\approx3^\circ$ over the last century which clearly contradicts the observations (Borkovits et al.~\cite{borkovitsetal04}). In order to resolve this inconsistency, Csizmadia et al.~\cite{csizmadiaetal09} carried out a combined optical and radio interferometric study to determine more precisely the spatial orientation of the inner orbital plane.  They used the CHARA Array, a six-element optical/IR interferometer located on Mount Wilson (ten Brummelaar et al.~\cite{tenbrummelaaretal05}), and e-EVN (European VLBI Network). The refined value of the mutual inclination in the Algol system was found to be $i_\mathrm{mut}=95^\circ\pm3^\circ$. While this value differs only by $5^\circ$ from the previously accepted 100$^\circ$, this small difference reduces by half the speed of the orbital plane precession, which tends to infinity for exactly perpendicular orbits (see Eq.~\ref{eq:Prec}). Therefore, this much slower precession rate is now in accord with the absence of any EDVs from the start of the precise scientific observations of Algol's eclipses. Note, moreover, parallel to this work, Zavala et al.~\cite{zavalaetal10} resolved all three components of Algol at optical wavelengths with NPOI. They resolved some orientation ambiguities that were present in the former works and obtained $i_\mathrm{mut}=86^\circ\pm5^\circ$.


We have left the space-based astrometric missions, Hipparcos (Perryman et al.~\cite{perrymanetal997}) and Gaia (Gaia collaboration~\cite{gaia16a}) to the end of this review. In regard to their primary missions, i.e., obtaining precise trigonometric parallaxes and, moreover especially in the case of Gaia, an almost complete three-dimensional map of our Milky Way, these have been hugely successful. In addition, any discrepancy between the short-term proper motion vectors measured with Hipparcos and with Gaia, compared to the long-term proper motion determined from the precise astrometric positions between the two missions, leads to the so-called `proper motion anomaly'.  Detections of this effect have already led to the identification of thousands of unresolved candidate binary and multiple star systems even down to substellar mass companions (Kervella et al.~\cite{kervellaetal19,kervellaetal22}; Brandt~\cite{brandt18,brandt21}). Moreover, Gaia, by itself, is even beginning to reveal the presence of multiple stars within many of the stellar images via anomalously large systemic errors in their astrometric solutions (see also, e.g, Kervella et al.~\cite{kervellaetal22}). From a wider perspective, these surveys, especially Gaia, may offer significant advances in the study of CHTs. Gaia is continuously gathering not only astrometric measurements of unprecedented precision for more than a billion stars, but simultaneously, high-precision photometric and high-resolution spectroscopic observations of even quite faint stars. Though these data have not been released yet, in the near future they should provide an inexhaustible supply of sources for statistical studies of EBs and, therefore, of (E)CHT-s as well. 

\section{Discussion and Conclusions}
\label{sec:discussion}

In this review we have dealt with a narrow, but important, subgroup of hierarchical multiple stellar systems---the most compact triple or quadruple stars. More specifically, we have focused on those members of this group which contain at least one eclipsing pair. The question naturally arises: Is there any reason to highlight such a very specific subset of the zoo of the multiple stellar systems? Is there anything specific that ECHTs can add to the science of EBs and, on the other hand, to the science of the multiple stars?

Addressing this question from an observational perspective, ECHTs are sources of an enormous amount of geometrical, dynamical, and astrophysical data that, in an optimal case, can be extracted with careful, sophisticated analyses of the available observational material.  As an illustration, we display in Table~\ref{Tab:RV-lc} an extended version of the well-known tabulation of the kinds of stellar and orbital parameters that can be obtained from spectroscopic and photometric observations of an EB (and their combinations). The original table is the grey shaded area in the upper left quarter of the extended version.

\begin{specialtable} 
\tiny
 \caption{System parameters that can be deduced from light curve and RV curve measurements}
 \label{Tab:RV-lc}
 \begin{tabular}{@{}l|ccccc|cccccc}
  \toprule
   & 1 & 2 & 3 & 4 & 5 & 6 & 7 & 8 & 9 & 10 & 11\\
  \midrule
$a_\mathrm{A}\sin i_\mathrm{in}$ or $a_\mathrm{B}\sin i_\mathrm{in}$ &\cellcolor[HTML]{C0C0C0} &\cellcolor[HTML]{C0C0C0}$\sqrt{}$&\cellcolor[HTML]{C0C0C0} &\cellcolor[HTML]{C0C0C0}$\sqrt{}$&\cellcolor[HTML]{C0C0C0} & &&&&  \\
$a_\mathrm{in}\sin i_\mathrm{in}$, $a_\mathrm{A,B}\sin i_\mathrm{in}$, $\mathcal{M}_\mathrm{A,B}\sin^3i_\mathrm{in}$&\cellcolor[HTML]{C0C0C0}&\cellcolor[HTML]{C0C0C0}&\cellcolor[HTML]{C0C0C0}$\sqrt{}$&\cellcolor[HTML]{C0C0C0}&\cellcolor[HTML]{C0C0C0}$\sqrt{}$ && &&&$\sqrt{}$&$\sqrt{}$\\
$a_\mathrm{in}$, $a_\mathrm{A,B}$, $\mathcal{M}_\mathrm{A,B}$, $\mathcal{R}_\mathrm{A,B}$, $\mathcal{L}_\mathrm{A,B}$, $d$ &\cellcolor[HTML]{C0C0C0}&\cellcolor[HTML]{C0C0C0}&\cellcolor[HTML]{C0C0C0}&\cellcolor[HTML]{C0C0C0}$(\sqrt{})$&\cellcolor[HTML]{C0C0C0}$\sqrt{}$ && &&&$\sqrt{}$&$\sqrt{}$\\ 
$e_\mathrm{in}$, $\omega_\mathrm{in}$, $P_\mathrm{in}$&\cellcolor[HTML]{C0C0C0}$\sqrt{}$&\cellcolor[HTML]{C0C0C0}$\sqrt{}$&\cellcolor[HTML]{C0C0C0}$\sqrt{}$&\cellcolor[HTML]{C0C0C0}$\sqrt{}$&\cellcolor[HTML]{C0C0C0}$\sqrt{}$ &&$\sqrt{}$&$\sqrt{}$ & & $\sqrt{}$ & $\sqrt{}$\\ 
$\gamma$& \cellcolor[HTML]{C0C0C0}&\cellcolor[HTML]{C0C0C0}$\sqrt{}$&\cellcolor[HTML]{C0C0C0}$\sqrt{}$&\cellcolor[HTML]{C0C0C0} &\cellcolor[HTML]{C0C0C0}$\sqrt{}$ & &&& $\sqrt{}$ & $\sqrt{}$ & $\sqrt{}$\\ 
$q_\mathrm{in}^\mathrm{sp}$&\cellcolor[HTML]{C0C0C0} &\cellcolor[HTML]{C0C0C0} &\cellcolor[HTML]{C0C0C0}$\sqrt{}$&\cellcolor[HTML]{C0C0C0} &\cellcolor[HTML]{C0C0C0}$\sqrt{}$ &&&&& $\sqrt{}$ & $\sqrt{}$\\ 
$q_\mathrm{in}^\mathrm{ph}$&\cellcolor[HTML]{C0C0C0}$(\sqrt{})$&\cellcolor[HTML]{C0C0C0} &\cellcolor[HTML]{C0C0C0} &\cellcolor[HTML]{C0C0C0}$(\sqrt{})$&\cellcolor[HTML]{C0C0C0}$(\sqrt{})$ && & ($\sqrt{}$) && ($\sqrt{}$) & ($\sqrt{}$)\\
$i_\mathrm{in}$, $\mathcal{R}_{A,B}/a_\mathrm{in}$, $\mathcal{L}_B/\mathcal{L}_A$, $\beta_{A,B}$, $A_{A,B}$, $F_{A,B}$, $x_{A,B}$, $\ell_3$&\cellcolor[HTML]{C0C0C0}$\sqrt{}$&\cellcolor[HTML]{C0C0C0} &\cellcolor[HTML]{C0C0C0} &\cellcolor[HTML]{C0C0C0}$\sqrt{}$&\cellcolor[HTML]{C0C0C0}$\sqrt{}$ &&&$\sqrt{}$ & & $\sqrt{}$ & $\sqrt{}$\\
$T_B(/T_A)$&\cellcolor[HTML]{C0C0C0}$\sqrt{}$&\cellcolor[HTML]{C0C0C0}(?)&\cellcolor[HTML]{C0C0C0}(?)&\cellcolor[HTML]{C0C0C0}$\sqrt{}$&\cellcolor[HTML]{C0C0C0}$\sqrt{}$ &&&$\sqrt{}$ && $\sqrt{}$ & $\sqrt{}$\\
\midrule
$a_{AB}\sin i_\mathrm{out}$ & & ($\sqrt{}$) & ($\sqrt{}$) & ($\sqrt{}$) & ($\sqrt{}$) & $\sqrt{}$ & &&& $\sqrt{}$ & $\sqrt{}$\\
$a_{C}\sin i_\mathrm{out}$ & &  &  &  &  & && & $\sqrt{}$ & $\sqrt{}$ & $\sqrt{}$\\
$a_\mathrm{out}/a_\mathrm{in}$ && & &&&&  $\sqrt{}$ & $\sqrt{}$ & & $\sqrt{}$ & $\sqrt{}$\\
$q_\mathrm{in}^\mathrm{dyn}$&& & &&&&  ($\sqrt{}$) &&&& ($\sqrt{}$)\\
$q_\mathrm{out}^\mathrm{dyn}$&& & &&&&  $\sqrt{}$ & $\sqrt{}$ &&& $\sqrt{}$\\
$q_\mathrm{out}^\mathrm{sp}$&& & &&&&&&& $\sqrt{}$ & ($\sqrt{}$)\\
$e_\mathrm{out}$, $\omega_\mathrm{out}$, $P_\mathrm{out}$ & & ($\sqrt{}$) & ($\sqrt{}$) & ($\sqrt{}$) & ($\sqrt{}$) & $\sqrt{}$ & $\sqrt{}$& $\sqrt{}$ & $\sqrt{}$ & $\sqrt{}$ & $\sqrt{}$\\ 
$\omega^\mathrm{dyn}_\mathrm{in}$, $\omega^\mathrm{dyn}_\mathrm{out}$, $\Omega^\mathrm{dyn}_\mathrm{in,out}$ && & &&&&  $\sqrt{}$ & ($\sqrt{}$) &&& $\sqrt{}$\\
$i_\mathrm{mut}$, $\Omega_\mathrm{in}-\Omega_\mathrm{out}$&& & &&&&  $\sqrt{}$ & $\sqrt{}$ &&& $\sqrt{}$\\
$i_\mathrm{in}$ ,$i_\mathrm{out}$&& & &&&&  ($\sqrt{}$) & $\sqrt{}$ && $\sqrt{}$ & $\sqrt{}$\\
$\mathcal{R}_\mathrm{A,B,C}$, $\mathcal{L}_\mathrm{C}/\mathcal{L}_\mathrm{A}$, $T_\mathrm{C}(/T_\mathrm{A})$, $\beta_\mathrm{C}$, $A_\mathrm{C}$, $F_\mathrm{C}$, $x_\mathrm{C}$, $\ell_4$& & & & &  &&&$\sqrt{}$ &&& $\sqrt{}$\\
\midrule
$\mathcal{M}_\mathrm{A,B,C}$, $\mathcal{R}_\mathrm{A,B,C}$, $T_\mathrm{A,B,C}$, $\mathcal{L}_\mathrm{A,B,C}$, $\beta_\mathrm{A,B,C}$, $A_\mathrm{A,B,C}$, $F_\mathrm{A,B,C}$, $x_\mathrm{A,B,C}$, $\ell_4$ &&&&&&&&&&&$\sqrt{}$ \\
$[M/H]$, $age$, $d$ &&&&&&&&&&& $\sqrt{}$\\
\bottomrule
\end{tabular}

\small
$1 =$ at least one photometric lightcurve; \\
$2 =$ one radial velocity curve (SB1); \\
$3 =$ two radial velocity curves (SB2); \\
$4 =$ at least one photometric lightcurve plus one RV curve (SB1); \\
$5 =$ at least one photometric lightcurve plus two RV curves (SB2); \\
$6 =$ LTTE dominated ETV; \\
$7 =$ ETV with dynamical effects; \\
$8= $ lightcurve with third body eclipses; \\
$9= $ RV curve of the tertiary; \\
$10=$ at least one photometric lightcurve plus three RV curves (SB3); \\
$11=$ Combination of lightcurve + ETV + RV (if any!) + SED + isochrones ({\sc Lightcurvefactory}); \\
\end{specialtable}

Closing our overview, we now discuss some aspects of the usefulness and significance of the extra information content that has become available with the analysis of ECHTs, in contrast to binary stars and/or wider multiple systems.

\noindent
{\bf Determination of the full spatial configurations of hierarchical triples and quadruples.} As was mentioned above, the mutual inclination(s) of the orbital planes is (are) perhaps the most important key parameter(s) in a hierarchical triple (multiple) system both from a dynamical and evolutionary point of view. It is very probably a primary tracer of the formation mechnism(s) of the given system (e.g., Tokovinin~\cite{tokovinin993}, Sterzik \& Tokovinin~\cite{sterziktokovinin02}, Fabrycky \& Tremaine~\cite{fabryckytremaine07}, Moe \& Kratter~\cite{moekratter18}, Tokovinin~\cite{tokovinin21}). However, the measurement of this parameter is quite challenging. As mentioned above in Sect.~\ref{sec:astrometry}, it can be readily calculated if one knows the observable inclinations ($i_\mathrm{in,out}$ and nodes ($\Omega_\mathrm{in,out}$) of both orbits. Since the inner pair is an EB, its inclination can be determined relatively easily from a basic analysis of the orbital light curve, at least to within an ambiguity of ($i$, $180^\circ-i$). Moreover, if such an ECHT happens to be a triple-lined (SB3) spectroscopic triple, its outer inclination may also be inferred from the amplitudes of the inner and outer RV curves, again, within an ambiguity of ($i$, $180^\circ-i$). For the determination of the two nodes, and also for $i_\mathrm{out}$ in non-SB3 cases, however, one generally needs astrometry. Since in CHTs the outer orbital separations in most cases are also at the subarcsec level, one usually needs some interferometric methods (at least speckle interferometry). This significantly reduces the number of the available CHTs for which the outer orbital elements can be determined. The most problematic part, however, is the measurement of the node of the inner orbit ($\Omega_\mathrm{in}$) which requires spatially resolving the inner orbit, where the size of CHTs is typically on the mas or sub-mas levels. Therefore, as far as we know, Algol is the only (E)CHT for which the mutual inclination has been determined in such a geometric, or `traditional', way.\footnote{Note also, for wider systems where astrometric measurements for both the inner and outer orbits are achievable, another problem arises, namely, insufficient coverage of the very long-period outer  orbits. Consequently, the number of wider multiples with robustly determined mutual inclinations is also modest (see, e.g. Tokovinin \& Latham~\cite{tokovininlatham17}).}

In contrast to the many caveats of such astrometry-based measurements, third-body perturbations in a tight ECHT offer a fast, direct, dynamical determination of this quantity.  Moreover, in contrast to the targets that can be resolved with interferometry, the collection of available systems is not limited to nearby and bright targets. In principle, the only data that are needed for the dynamical measurement of the mutual inclination in a tight ECHT, are single-band, high-quality photometric observations of the eclipses of the inner EB over a few, or at least {\it one}, period of the outer orbit. Then, once the mid-eclipse times have been found, an analysis of the DE dominated ETVs of the given system can lead directly to $i_\mathrm{mut}$, as one of the fitting parameters. In such a way, Borkovits et al.~\cite{borkovitsetal16} determined the mutual inclinations of more than 40 tight triple stars (most of them ECHTs) in the \textit{Kepler} sample (see also Sect.~\ref{sec:ETVs}). Moreover, in addition to the relative orientation of the orbits, DE dominated ETVs can also be used to determine other dynamically important parameters, such as the dynamical arguments of periastron ($\omega^\mathrm{dyn}_\mathrm{in,out}$) which directly affect the ZKL cycles and, therefore, the dynamical evolution of the CHTs (e.~g., Naoz~\cite{naoz16}).

We also note that ETV analysis uses only a single extracted piece of the information content that is encoded into an EB light curve. Extending the analysis to the complete light curve, and including the available information about the eclipse depths and/or duration variations (or their absence), the mutual inclination and other dynamical parameters can be obtained with an even higher precision.

In addition to the dynamical determination of the relative orbital inclinations, the discovery of triply eclipsing systems offers another opportunity since the shape and timing of the extra eclipses are extremely sensitive to the relative configuration and movement of the three stars (or, two stars and a circumbinary planet).  Via careful modelling of the extra eclipses, the mutual inclination can be accurately determined even in the absence of detectable third-body perturbations (Borkovits et al.~\cite{borkovitsetal13}; Masuda et al.~\cite{masudaetal15}; Mitnyan et al.~\cite{mitnyanetal20}; Borkovits et al.~\cite{borkovitsetal22}).

\noindent
{\bf Astrophysical relevance of the combined analyses of all available data on ECHTs.} The important role of EBs and their careful investigations, viz., `the Royal Road to Stellar Astrophysics' (Batten~\cite{batten05}) in the history of the modern stellar astrophysics, is widely acknowledged and not necessary to repeat here.\footnote{For a very conscise summary see e.~g. Southworth~\cite{southworth20} and references therein.} Currently the masses and radii of some of the brightest detached EBs are known with accuracies well below 1\% (see, e.g., Maxted et al.~\cite{maxtedetal20}; Southworth~\cite{southworth21a,southworth21b}), and some hundreds of additional EBs are characterized with an accuracy of 2\%, or better (Southworth~\cite{southworth15}). These EBs can then be used for the refinement of stellar evolutionary models and to resolve some recent existing problems, such as, e.g., the radius anomaly (see, e.g. Garrido et al.~\cite{garridoetal19}) of low-mass stars.  The question arises as to how ECHTs relate to these issues. In some studies the presence of a third body is considered only as a source of  unwanted additional noise which is to be eliminated. However, it is evident that in the case of an inappropriate treatment, the effects of any additional components may produce not only larger uncertainties, but also systematic errors. The solution should be the integration of all the observable third-body effects (eclipse timing, depth, duration variations, extra eclipses, extra flux, systemic RV variations, RVs of the additional bodies etc.) into a complex spectro-photodynamical analysis (see, e.g., Ragozzine \& Holman~\cite{ragozzineholman10}, Carter et al.~\cite{carteretal11}, P\'al~\cite{pal12}, Bro\v{z}~\cite{broz17}, Short et al.~\cite{shortetal18}, Borkovits et al.~\cite{borkovitsetal19c}).  Moreover, a promising further direction is the inclusion of the modelling of the net stellar energy distribution (SED), which enables the determination of temperatures for each stellar component with precision close to or better than 1\%. Recently, for example, Miller et al.~\cite{milleretal20} have developed a sophisticated method for the SED modelling of totally eclipsing detached EBs, and demonstrated that under ideal circumstances their method is able to reach the required accuracy in the two stars' temperatures. Their method, however, cannot be adopted directly to multiple stellar systems, but it does indicate a promising direction. On the other hand, in this regard, we also refer to Csizmadia~\cite{csizmadia20} who pointed out that stellar effective temperatures are affected by  stellar spots, especially at subsolar masses and, therefore, the SED-fitted stellar temperatures can be different from the spectroscopically derived ones because of the spots. 

Amongst the most recent efforts to obtain very precise stellar and dynamical data with an analysis of ECHTs we refer again the earlier mentioned work of Yenawine et al.~\cite{yenawineetal21} on the triply eclipsing triple KOI-126. They were able to obtain accuracies better than 1\% for the masses and radii despite the fact that RV data were available only for the outer (most massive and brighter) component. Note also the still unpublished results of Gaulme et al.~\cite{gaulmeetal22} who obtained 1--2\% accuracy in the masses of KIC~7955301, similar to KOI-126, with the use of RV data for only the outer component.  But, in contrast to KOI-126, there are no third-body eclipses which would offer even stronger dynamical and geometric constrains for the fundamental parameters.

{\bf Investigation of small, higher order effects.} Accurate modeling of such higher order atmospheric effects such as limb darkening and gravitational brightening are especially important for the precise analysis of transiting extrasolar planets (see, e.g. Csizmadia et al.~\cite{csizmadiaetal13}).  And, these may also be important for future extremely high precision analyses of EBs. In this regard, inclination varying EBs (Sect.~\ref{sec:depthvar}) may be used for scanning the distribution of the emitted flux over the entire stellar disc.

Finally, we fully anticipate significant advances in these fields in the near future. For example, the individual astrometric (and, of course, RV and photometric) measurements of Gaia will become available for billions of stars, including ECHTs -- either known or yet to be discovered. These measurements may allow for the precise astrometric determinations of the outer orbits in some tight ECHTs having dynamical solutions. Adding these high precision astrometric data to the complex analyses one may expect unprecedented high quality results in many of the fundamental stellar and dynamical parameters.

\dataavailability{The data in Figs.~\ref{fig:3beclipses} and ~\ref{fig:EDVs} were provided by NASA and obtained either from the Kepler Eclipsing Binary Catalog at \url{http://keplerebs.villanova.edu/} or the Mikulski Archive for Space Telescopes (MAST) at \url{https://mast.stsci.edu/portal/Mashup/Clients/Mast/Portal.html}.}

\acknowledgments{The author is pleased to thank Saul Rappaport for his valuable comments and linguistic editions. The anonymous referees are also acknowledged for their quick and positive reports. This paper includes data collected by the \textit{Kepler} and \textit{TESS} missions and obtained from the MAST data archive at the Space Telescope Science Institute (STScI). Funding for the Kepler mission is provided by the NASA Science Mission Directorate. Funding for the TESS mission is provided by the NASA Explorer Program. STScI is operated by the Association of Universities for Research in Astronomy, Inc., under NASA contract NAS5--26555. We used the Simbad service operated by the Centre des Donn\'ees Stellaires (Strasbourg, France) and the ESO Science Archive Facility services (data obtained under request number 396301). This research has made use of NASA's Astrophysics Data System.}

\conflictsofinterest{The author declares no conflict of interest.} 


\end{paracol}
\reftitle{References}

\end{document}